\documentclass[11pt,a4paper]{article}
\usepackage{jheppub}
\usepackage{xcolor}
\usepackage{amsmath,amssymb,amsfonts}
\usepackage{physics}
\usepackage{graphicx}
\usepackage{svg}
\graphicspath{{./images}}
\usepackage{subcaption}
\usepackage{array}
\usepackage{tikz}
\usepackage{url}
\usepackage{multirow}
\usepackage{comment}
\usepackage{tabularx}
\newcolumntype{Y}{>{\centering\arraybackslash}X}
\usepackage{makecell}
\newcommand{\eq}[1]{eq.~(\ref{#1})}

\def\blue{\color{blue}}

\usepackage{hyperref}

\usepackage{hyperref}

\def\blue{\color{blue}}

\def\stw{s_{\theta_W}}
\def\ctw{c_{\theta_W}}
\def\ttw{t_{\theta_W}}
\def\lra#1{\overset{\text{\scriptsize$\leftrightarrow$}}{#1}}

\newcommand\bm[1]{\boldsymbol{#1}}

\begin{document}

\definecolor{lime}{HTML}{A6CE39}
\DeclareRobustCommand{\orcidicon}{\hspace{-1mm}
	\begin{tikzpicture}
	\draw[lime, fill=lime] (0,0) 
	circle [radius=0.12] 
	node[white] {{\fontfamily{qag}\selectfont \tiny \,ID}};
	\draw[white, fill=white] (-0.0525,0.095) 
	circle [radius=0.007];
	\end{tikzpicture}
	\hspace{-3mm}
}

\foreach \x in {A, ..., Z}{\expandafter\xdef\csname orcid\x\endcsname{\noexpand\href{https://orcid.org/\csname orcidauthor\x\endcsname}
		{\noexpand\orcidicon}}
}

\newcommand{\orcidauthorA}{0009-0000-2346-2273}
\newcommand{\orcidauthorB}{0000-0001-5815-4182}
\newcommand{\orcidauthorC}{0000-0003-4398-4698}

\title{Towards the HEFT-hedron: the complete set of positivity constraints at NLO}
\preprint{TIFR/TH/24-26}

\author[a]{Debsubhra Chakraborty\orcidB{}}
\author[a]{Susobhan Chattopadhyay\orcidA{}}
\author[a]{Rick S. Gupta\orcidC{}}

\affiliation[a]{Tata Institute of Fundamental Research, Homi Bhabha Road, Colaba, Mumbai 400005, India}

\emailAdd{debsubhra.chakraborty@tifr.res.in }
\emailAdd{susobhan.chattopadhyay@tifr.res.in}
\emailAdd{rsgupta@theory.tifr.res.in}

\begin{abstract}
{We present the complete set of positivity bounds on the Higgs Effective Field Theory (HEFT)  at next-to-leading order (NLO). We identify the 15 operators that can be constrained by positivity, as they contribute to $s^2$-growth in the amplitude for longitudinal gauge-Higgs scattering, that is to all possible 2-to-2 scattering processes involving longitudinal gauge bosons,  $V_L = W_L^\pm, Z_L$, and the  Higgs boson, $h$. We find two sets of constraints: (i) specific linear combinations of CP-even Wilson coefficients (WCs) must be positive, and (ii) the magnitudes of some WCs---including all CP-odd ones---must be smaller than products of other CP-even WCs. We  present our final constraints on the 15 dimensional HEFT space and show how known positivity bounds on the 3 dimensional  space of dimension 8 SMEFT  can be recovered from them. We find that  only about $5\%$ of the parameter space for WCs of HEFT operators at NLO complies with these positivity constraints.  Additionally, we obtain double-sided bounds on these WCs by fully exploiting the implications of  unitarity and $st$-crossing symmetry. For WCs contributing to the vector boson scattering process our final constraints are in most cases significantly stronger than the experimental ones. For the   $V_L V_L, hh \to hh$ and $V_LV_L, hh \to V_Lh$ process, there are no reported experimental limits and our theoretical constraints provide the first  bounds.  }
\end{abstract}
\maketitle
\par\noindent\rule{\textwidth}{0.4pt}

\setcounter{page}{0}
\pagenumbering{arabic}

\section{Introduction}

The discovery of the Higgs boson~\cite{ATLAShiggs, CMShiggs}  was a landmark event in the history of particle physics. A detailed characterisation of the  Higgs and electroweak sector is arguably the most concrete goal of particle physics research today. In the absence  of any direct signs of new physics at the Large Hadron Collider (LHC), effective field theories (EFT) provide a natural framework to parameterise deviations in Higgs and electroweak physics from   Standard Model (SM) predictions.

The Standard Model Effective Field Theory (SMEFT)---that extends the Standard Model (SM) lagrangian by a series of higher dimensional operators---has thus become the standard way to parameterise deviations in indirect searches for both experimentalists and theorists.  This has led to a sophisticated program of devising experimental strategies to optimally and maximally probe the space of the Wilson coefficients (WCs) of SMEFT operators. An important   development, complementing this experimental program, has been the use of theoretical principles like causality/analyticity, locality, unitarity and crossing symmetry to provide rigorous  bounds on this space of EFT coefficients.  It was   shown  in Ref.~\cite{Allan_Adams_2006}\footnote{An earlier application of similar ideas to the chiral lagrangian was made in Ref.~\cite{Pham, Anant, Pennington}.} that the above theoretical principles enforce the positivity of the WCs that give    $s^{2n}$  growth in the forward amplitude for 2$\to$2   scattering of goldstones and photons,   $n \geq 1$ being an integer.  A lot of recent work has focused on further extending these arguments to maximally constrain the space of WCs. This has culminated in the derivation of the full set of constraints from 2$\to$2  scattering on the space of EFT WCs  for causal and unitary theories~\cite{moments,Tolley,Extremal, Arkani-Hamed,Sinha}--- thus giving rise to the so-called `EFT-hedron'~\cite{Arkani-Hamed}, the volume within which the WCs must lie.

In the SMEFT this approach  has led to positivity bounds on the coefficient of the $s^2$ term in the forward amplitude which in turn can be translated to bounds on the WCs of operators at the dimension-8 (D8) level---the lowest  order at  which an $s^2$ growth in amplitudes becomes possible~\cite{briva, zhang1, zhang2, remmen1, remmen2, Ghosh1}.\footnote{In fact,  the positivity bounds, in general,  constrain  a sum of dimension-8 WCs and a quadratic function of dimension-6 WCs.  Nevertheless, it has been possible to individually constrain the  dimension-8  contribution~\cite{zhang1,zhang2, remmen1}  as we  discuss briefly in Sec.~\ref{smeftsec}.} It has been recently shown that the more optimal application of unitarity and crossing symmetry developed in Ref.~\cite{Tolley, Extremal},   can  complement these  constraints by providing   double sided bounds on   WCs~\cite{Chen:2023bhu}.  The phenomenological relevance of the SMEFT positivity bounds is, however, somewhat  limited because the leading  deviations from the SM are parameterised by dimension-6 operators that are not subject to positivity bounds.

As is increasingly being recognised, however,  the SMEFT is the not the most general framework to characterise indirect effects. The space of SMEFT WCs at a given order, is in fact a subspace of the more general Higgs EFT (HEFT). This is because of the implicit assumptions made while writing the SMEFT lagrangian. First, unlike the SMEFT, in the HEFT it is not  assumed that the observed Higgs boson, $h$, is part of the electroweak doublet that breaks electroweak symmetry. Furthermore,  even in theories where $h$ is part of a electroweak symmetry breaking  doublet,  if the states  being integrated out obtain a large fraction of their mass from the electroweak VEV, the resulting low energy theory is the HEFT and not the SMEFT~\cite{rattazzi, craig1, craig2}. 

In this work we obtain the complete set of positivity bounds on HEFT lagrangian at  the next to leading order (NLO). The bounds apply to WCs of 15 operators in the NLO lagrangian with 4 derivatives  that contribute to scattering of Higgs and longitudinal gauge bosons, i.e. to the  $V_L V_L  \to V_L V_L$, $hh$,  $hh \to V_L h,~{hh}$   and $V_L h \to V_Lh$ processes processes, where $V_L = W_L^\pm, Z_L$ are the longitudinal components of the electroweak bosons and $h$ is the observed Higgs boson. No other process has a forward amplitude growing as $s^2$ or faster in the HEFT NLO lagrangian.   First,  we consider elastic scattering between superpositions of gauge and Higgs bosons  to obtain analytical positivity constraints. These result in an allowed region that forms a convex cone~\cite{cone} occupying about ${5\%}$ of  the HEFT parameter space. We then use the methods  developed in Refs.~\cite{Tolley, Extremal, Chen:2023bhu} to numerically derive double-sided bounds to cap this conical region. While ours is the first work to derive positivity constraints for the HEFT, there have been previous works that considered vector boson scattering to derive some of the  positivity constraints on the higgsless electroweak chiral lagrangian~\cite{ewcl1, ewcl2}. Even for this case,   we go beyond the work of Refs.~\cite{ewcl1, ewcl2}---first,   by deriving the complete set of analytical positivity constraints on the WCs involved, and  then by  imposing  double sided bounds  on them.

The positivity bounds for HEFT are phenomenologically more  relevant compared to  the SMEFT case, because they appear already at the next leading order. It is also worth emphasizing that  being the most general lagrangian at the weak scale, it is the  HEFT, and not the SMEFT, that provides the most general parametrization of the coefficient of the  $s^2$ term in the forward amplitude for  longitudinal gauge-Higgs scattering. We show this explicitly in the next section. Existing positivity bounds on   SMEFT constrain a set of 3  dimension-8 operators. We will show how these known SMEFT results can be recovered  by   obtaining the intersection of  the three-dimensional SMEFT  hyperplane    with the NLO HEFT-hedron, the 15-dimensional   region  allowed by our positivity bounds.

This paper is organized as follows. In Sec.~\ref{two}, we parameterize longitudinal gauge-Higgs scattering in the SMEFT, HEFT and  by imposing only $U(1)_{em}$. In Sec.~\ref{three} we analytically derive  positivity constraints that result in the aforementioned conical allowed region in the space of HEFT WCs. We then use the methods of Ref.~\cite{Tolley, Extremal, Chen:2023bhu}   to numerically obtain   double-sided bounds to  cap this positivity cone in Sec.~\ref{dsided}.  In Sec.~\ref{hefthedron}  we show our final results and compare them with existing experimental bounds. Finally, we provide concluding remarks in Sec.~\ref{conclusions}.

\paragraph{Note added:}   The main results of this work were already presented in the general meeting of the LHC EFT working group  on December 2, 2024~\cite{talk}. As we were in the final stages of preparing this manuscript, a preprint~\cite{Rodd} appeared on the arxiv that has some overlap with our results. Working on the set of 5 custodial invariant HEFT operators at NLO  that receive positivity bounds, Ref.~\cite{Rodd} derives the analytical positivity bounds mentioned above. Our results in Table~\ref{relations} are in complete agreement with those of Ref.~\cite{Rodd}. Ref.~\cite{Rodd}   also shows that if the HEFT region allowed by positivity  is projected onto the SMEFT plane it contains regions outside the region allowed by positivity in SMEFT. They argue that observation of new physics  outside the SMEFT allowed region but within the HEFT allowed region, would  indicate that HEFT and not the SMEFT provides the correct low energy description. In Sec.~\ref{r2} we obtain similar results and provide our perspective on how and when positivity can indicate that the low energy theory is not SMEFT but the more general HEFT.   

\section{Parametrizing  longitudinal gauge-Higgs scattering} 
\label{two}

In this section we present the  amplitude for longitudinal gauge-Higgs scattering,  i.e. for the  $V_L V_L  \to V_L V_L$, $V_Lh$, $hh$,  $hh \to V_L h,~{hh}$   and $V_L h \to V_Lh$ processes. We are interested in obtaining the part of the  forward amplitude that  grows as $s^2$ and is thus subject to positivity bounds. We will write the amplitude at tree-level using four different parametrizations. For first three parametrizations we will use the following lagrangians: (1) the HEFT NLO lagrangain, (2) a $U(1)_{em}$ parametrization using anomalous couplings and (3) the SMEFT lagrangian up to the dimension-8 level. Finally, assuming only $U(1)_{em}$ and crossing symmetry, we will provide a general parametrization of the amplitude in terms of Mandelstam invariants and show that there is a one-to-one mapping between the free parameters in this approach and the lagrangian couplings of the first two  lagrangian frameworks. In particular we will provide the explicit mapping between these three parametrizations that  will make the interpretation of our final bounds on HEFT WCs in terms of anomalous couplings straightforward. The number of parameters in the SMEFT parametrization, however, is smaller than that in the other cases. This implies  constraints on the space of HEFT WCs/anomalous couplings if they arise from SMEFT at dimension $8$.

\subsection{HEFT parametrization}\label{HEFT_param}
We begin with  the HEFT  where it is especially straightforward to identify pieces of the gauge Higgs-scattering forward amplitudes that grow as $s^2$. In  the HEFT lagrangian the observed Higgs boson, $h$,  transforms as a singlet under electroweak symmetry whereas the three goldstone modes  associated to the breaking  of electroweak symmetry transform non-linearly under it. Here we will follow the formalism in Ref.~\cite{longhitano1, longhitano2, applequist1, applequist2,Buchalla1, Alonso1, Buchalla2, Brivio1, Brivio2, Sun}  where the spontaneous symmetry breaking pattern is assumed to be,\footnote{See Ref.~\cite{georgi, murayama}  for an alternative formulation with the spontaneous symmetry breaking pattern, $SU(2)_L\times U(1)_Y\to U(1)_{em}$.}
\begin{equation}
SU(2)_L\times SU(2)_R\to SU(2)_C,    
\end{equation}
$SU(2)_R$ and $SU(2)_C$ being  approximate symmetries. The resulting goldstones, $\phi_I$,  reside in the  matrix, $U = \exp{\left[\frac{i}{v}\phi_{I}\sigma^{I}\right]}$,  where, $I\in{1,2,3}$ and $v=246$ GeV is the electroweak VEV. The transformation of $U$  under the  group $SU(2)_L\otimes SU(2)_R$ is given by, 
\begin{eqnarray}
U(x) \to LU(x)R^{\dagger}.
\end{eqnarray}
and the covariant derivative  of $U$ is defined as, $D_{\mu}U=\partial_{\mu}+i\hat{W}_{\mu}U-iU\hat{B}_{\mu}$ where,
\begin{eqnarray}
\hat{W}_{\mu}=g\frac{\vec{\sigma}}{2}\vec{W}_{\mu};\hspace{1.5cm} \hat{B}_{\mu}=g^{\prime}\frac{\sigma_3}{2}B_{\mu}.
\end{eqnarray}
The building blocks of the HEFT lagrangian are,
\begin{eqnarray}
\mathbf{V}_{\mu}=iUD_{\mu}U^{\dagger} \hspace{1.5cm}\mathbf{T}=U\frac{\sigma_3}{2}U^{\dagger}
\end{eqnarray}
and the observed Higgs $h$, which transform   as follows,
\begin{eqnarray}
\mathbf{V}_{\mu} \to L\mathbf{V}_{\mu}L^{\dagger}\hspace{1cm} \mathbf{T} \to L\mathbf{T}L^{\dagger}\hspace{1cm} h \to h.
\end{eqnarray}
The presence of a $\sigma_3$ in the definition of {$\mathbf{V}_{\mu}$ and} $\mathbf{T}$   explicitly breaks $SU(2)_R\to U_Y(1)$ in the HEFT lagrangian (see for eg. the discussion in Ref.~\cite{murayama}).

The list of NLO operators in the HEFT,  presented in different 
works (see  Ref.~\cite{Alonso1, Buchalla2, Brivio1, Brivio2, Sun}) differ  from each other based on the   criteria used by the authors in organising the HEFT expansion.  In all these lists, however,  the only operators, that can result in an $s^2$ growth in a $2 \to 2$ scattering process are those with 4 derivatives and no field strengths, i.e.  operators of the form  of the $UhD^4$. Note that $2 \to 2$ scattering amplitudes involving fermions and photons do not receive NLO HEFT contributions  that grow  faster than $s^{2}$.  Thus in order to obtain the complete set of positivity bounds on the HEFT at NLO, it is suffices to consider only  longitudinal gauge-Higgs scattering.

In Table~\ref{Table1}, we reproduce the  list all the $15$ operators of type $UhD^4$ presented in Ref.~\cite{Sun}. These operators contribute to the  amplitude for Higgs and goldstone scattering,  $\phi_i \phi_j \to \phi_k \phi_l$ with the indices taking values from 1--4 and $\phi_4\equiv h$; this amplitude  is identical to longitudinal gauge-Higgs scattering in the high energy limit  by  the goldstone boson equivalence theorem~\cite{gbe}. The functions, $\mathcal{F}^{UhD^4}_{i}(h)$, in Table~(\ref{Table1})  are given by,
\begin{equation}
\mathcal{F}^{UhD^4}_{i}(h)=1+\sum_{r=1}c_{ir}^{UhD^4}\left(\frac{h}{v}\right)^r\hspace{1cm}\nonumber
\end{equation}
for $i=1,\cdots 15$.  These operators appear in the NLO  HEFT lagarangain at NLO as follows, 
\begin{eqnarray}
\mathcal{L}_{HEFT}^{NLO}\supset\sum\limits_{n=1}^{15}c_i\mathcal{O}_i^{UD^4}+\cdots\label{HEFT_WCs}
\end{eqnarray}
where $c_i$ is the WC of the operator, $\mathcal{O}^{UhD^4}_{i}$. As far as the power-counting for these WCs is concerned, we assume, 
\begin{equation}\label{pc}
    c_i \sim v^2 \Lambda^2  \left(\frac{D}{\Lambda}\right)^{n_D} \left(\frac{h}{v}\right)^{n_h}.
\end{equation}
where $v/\Lambda \geq 1/4 \pi$. As we will soon discuss, if these HEFT operators arise from SMEFT, the  fact that $SU(2)_L \times U(1)_Y$ is then linearly realised,  will imply further suppression in some   linear combinations of the WCs.
\begin{table} [t]
\centering
\renewcommand{\arraystretch}{1.4} 
\begin{tabular}{@{} >{}c<{}  >{}c<{}  >{}c<{} >{}c<{}  >{}c<{}@{}} 
\hline\hline
\text{Process} & $i$   &    $\mathcal{O}^{UhD^4}_i$ & \text{CP} & $SU(2)_C$ \\ \hline
\multirow{5}{*}{$V_LV_L\to V_LV_L$}&$1$ & $\langle \mathbf{V}_\mu\mathbf{V}^{\mu}\rangle^2\mathcal{F}^{UhD^4}_1(h)$  & $+$& P\\
&$2$ & $\langle \mathbf{V}_\mu\mathbf{V}_{\nu}\rangle\langle\mathbf{V}^\mu\mathbf{V}^{\nu}\rangle\mathcal{F}^{UhD^4}_2(h)$ & $+$& P\\
& $3$ & $\langle \mathbf{T}\mathbf{V}_\mu\rangle\langle\mathbf{T}\mathbf{V}_\nu\rangle\langle \mathbf{V}^{\mu}\mathbf{V}^{\nu}\rangle\mathcal{F}^{UhD^4}_3(h)$ & $+$& V\\
&$4$ & $\langle \mathbf{T}\mathbf{V}_\mu\rangle\langle\mathbf{T}\mathbf{V}^\mu\rangle\langle \mathbf{V}^{\nu}\mathbf{V}_{\nu}\rangle\mathcal{F}^{UhD^4}_4(h)$  & $+$ & V\\
& $5$ & $(\langle \mathbf{T}\mathbf{V}_\mu\rangle\langle\mathbf{T}\mathbf{V}^\mu\rangle)^2\mathcal{F}^{UhD^4}_5(h)$  & $+$& V\\\hline
\multirow{4}{*}{$V_LV_L\to V_Lh$} & $6$ & $\langle \mathbf{V}_\mu \mathbf{V}^{\mu}\rangle\langle\mathbf{T}\mathbf{V}_\nu\rangle\frac{D^{\nu}h}{v}\mathcal{F}^{UhD^4}_6(h)$  & $-$& V\\
& $7$ & $\langle \mathbf{V}_{\mu}\mathbf{V}_{\nu} \rangle \langle \mathbf{T} \mathbf{V}^{\mu} \rangle \frac{D^{\nu}h}{v}\mathcal{F}^{UhD^4}_7(h)$ &  $-$ & V\\
& $8$ &  $i \langle \mathbf{T}\mathbf{V}_\mu\mathbf{V}_\nu\rangle \langle \mathbf{T}\mathbf{V}^{\mu}\rangle \frac{D^{\nu}h}{v}\mathcal{F}^{UhD^4}_8(h)$ & $+$& V  \\
& $9$ & $\langle \mathbf{T}\mathbf{V}_{\mu} \rangle \langle \mathbf{T} \mathbf{V}^{\mu} \rangle \langle \mathbf{T}\mathbf{V}_{\nu}\rangle\frac{D^{\nu}h}{v}\mathcal{F}^{UhD^4}_9(h)$ & $-$ & V\\\hline
\multirow{4}{*}{$V_LV_L\to hh$} &  $10$ & $\langle \mathbf{V}_{\mu}\mathbf{V}_{\nu}\rangle \frac{hD^{\mu}D^{\nu}h}{v^2}\mathcal{F}^{UhD^4}_{10}(h)$ & $+$& P\\
& $11$ & $\langle\mathbf{V}_{\mu}\mathbf{V}^{\mu}\rangle\frac{D_{\nu}hD^{\nu}h}{v^2}\mathcal{F}^{UhD^4}_{11}(h)$ & $+$& P\\
& $12$ & $\langle \mathbf{T}\mathbf{V}_{\mu} \rangle \langle \mathbf{T}\mathbf{V}_{\nu}\rangle\frac{hD^{\mu}D^{\nu}h}{v^2}\mathcal{F}^{UhD^4}_{12}(h)$ & $+$ & V\\
& $13$ & $\langle \mathbf{T}\mathbf{V}_{\mu}\rangle \langle \mathbf{T}\mathbf{V}^{\mu}\rangle \frac{D_{\nu}hD^{\nu}h}{v^2}\mathcal{F}^{UhD^4}_{13}(h)$ & $+$& V\\\hline
$V_L h\to hh$ & $14$ &  $\langle \mathbf{T}\mathbf{V}_{\mu}\rangle\frac{hD^{\nu}hD_{\nu}D^{\mu}h}{v^3}\mathcal{F}^{UhD^4}_{14}(h)$ & $-$& V \\ \hline
$hh \to hh$ & $15$ & $\frac{1}{v^4}h^2(D_{\mu}D_{\nu}h)(D^{\mu}D^{\nu}h)\mathcal{F}^{UhD^4}_{15}(h)$ & $+$& P \\ \hline\hline
\end{tabular}
\caption{The list of $15$ operators of type $UhD^4$ in the HEFT lagrangian at NLO in the basis of Ref.~\cite{Sun}. These are the complete set of operators at NLO that contribute to the $s^2$ piece in the forward amplitude for  2$\to$2 scattering involving goldstones and the Higgs boson---and thus to the the $s^2$ piece in the forward amplitude of longitudinal gauge-Higgs scattering. We also provide the CP properties of the operators in the fourth column, while the fifth column indicates whether each operator preserves (P) or violates (V) the custodial symmetry, $SU(2)_C$. The contractions of $SU(2)$ group indices between $\mathbf{T}$ and $\mathbf{V}_{\mu}$ are represented by the symbol $\langle \cdots \rangle$.  \label{Table1}}
\end{table}

Note that while presenting  the longitudinal gauge-Higgs scattering amplitudes, we will consider only  a single insertion of a HEFT NLO WC. If we consider more insertions, operators from other categories in Ref.~\cite{Sun}, for eg. $UhD^2X$, might also result in $s^2$ growth in the amplitude. These contributions will, however, be suppressed given the power counting scheme in \eq{pc} and can therefore be safely ignored.


\subsection{$U(1)_{em}$ parametrization  using anomalous couplings}

\label{anom}

We now present a $U(1)_{em}$  invariant  lagrangian containing all possible vertices  to parameterize longitudinal  gauge-higgs scattering. The couplings accompanying   individual $U(1)_{em}$  invariant lagrangian terms are   called anomalous couplings. These anomalous couplings  provide an especially convenient parametrization of deviations beyond the SM in experimental and phenomenological studies. 


 To obtain an amplitude that grows as  $s^2$, the anomalous vertices can be of the following forms: $\partial V V^2, V^4, \partial h V^3, (\partial h)^2 V^2, (\partial h)^3 V$ and $(\partial h)^4$. While the the $\partial V V^2$ terms contribute via tree-level exchange diagrams, the rest of the vertices are contact terms. That these terms can lead to an $s^2$ growth can be understood by recalling that longitudinally polarised vector bosons, $V_L^\mu=W^{\pm,\mu}_L, Z_L^\mu$, contribute a longitudinal polarization vector, $\epsilon^V_{L, \mu}\approx p_\mu/m_V$, to the amplitude. We  modify the SM lagrangian by adding terms of each of the above categories, 
 \begin{equation}
     {\cal L}_{BSM}\supset {\cal L}_{SM} +\Delta {\cal L}_{\partial V V^2}+\Delta {\cal L}_{V^4} +{\cal L}_{(\partial h)^2V^2} +{\cal L}_{\partial hV^3} +{\cal L}_{(\partial h)^3V} +{\cal L}_{(\partial h)^4}.
 \end{equation}
 As already mentioned in the previous section we ignore contributions to the amplitude involving two insertions of BSM couplings. 

Let us begin with vertices of the form $\partial V V^2$ formed out of 3 gauge bosons and a single derivative, the so-called  Triple Gauge couplings (TGC), that were first presented in Ref.~\cite{Hagiwara},
\begin{eqnarray}
    \Delta {\cal L}_{\partial V V^2} &=&  i g \ctw \left[
      {\delta g_1^Z} Z_\mu \left(W^-_\nu {\cal W}^{+\mu\nu} - W^+_\nu
         {\cal W}^{-\mu\nu}\right) 
      + \delta \kappa^Z W^-_\mu W^+_\nu Z^{\mu\nu}
      \right] \nonumber\\&& +i e~ {\delta \kappa^\gamma} W^-_\mu W^+_\nu A^{\mu\nu}-g_4 W^-_\mu W^+_\nu (\partial^\mu V^\nu+\partial^\nu V^\mu)+g_5 \epsilon^{\mu \nu \rho \sigma} W^+_{\mu} \lra{D}_\rho W^-_{\nu}Z_\sigma \nonumber\\&& +i g \ctw \delta \tilde{\kappa}^Z  W^-_\mu W^+_\nu \tilde{Z}^{\mu\nu}
      +i e~ {\delta \tilde{\kappa}^\gamma} W^-_\mu W^+_\nu \tilde{A}^{\mu\nu}.
\end{eqnarray}
where,
\begin{eqnarray}
   {\cal W}^{+\, \mu\nu}=D^e_\mu W^{+, \nu}-D^e_\nu W^{+, \mu},
\end{eqnarray}
$D^e_\mu= \partial_\mu +i e A_\mu$ being the covariant derivative.  In the above lagrangian the first three terms are even under both charge conjugation (C) and  parity (P), while all other  terms are either C-odd or  P-odd or both. As far as longitudinal vector boson scattering is concerned,   single insertions of only P and CP-even vertices can lead to an $s^2$ growth in the amplitude.  

The HEFT contributions to the three C and P-even TGCs do not arise from  the operators in Table~\ref{Table1},  but from the following three other operators from the NLO basis of  Ref.~\cite{Sun},   
\begin{eqnarray}\label{oblique}
\beta_0 \frac{v^2}{2}\langle \textbf{T} \textbf{V}_\mu\rangle \langle \textbf{T} \textbf{V}^\mu\rangle+\alpha_{33}  {g^2}\langle \textbf{T} \textbf{W}_{\mu\nu} \rangle \langle \textbf{T} \textbf{W}^{\mu\nu}\rangle
+\alpha_{WB} \frac{g g'}{2}\langle \textbf{T} \textbf{W}_{\mu\nu} \rangle B^{\mu\nu} 
\end{eqnarray}
where  $\textbf{W}_{\mu \nu}= W^a_{\mu \nu} \sigma^a/2$. 
Working  in the  input parameter scheme, $\{m_W, m_Z, \alpha_{em}\}$, we evaluate the contributions  from  the  operators in \eq{oblique} to the above anomalous couplings in the unitary gauge,  carefully taking into account input parameter redefinitions. This gives, 
\begin{eqnarray}
&& \delta g^{1}_{Z} =  
    \frac{1}{2 \stw^2}\frac{\delta m_Z^2}{m_Z^2}
  , \quad 
	\delta \kappa^{\gamma}=-\frac{g^2}{2}\alpha_{WB}-g^2 \alpha_{33}  \nonumber\\ && \delta \kappa^{Z}= \frac{1}{2 \stw^2}\frac{\delta m_Z^2}{m_Z^2}+\frac{g^2}{2}  \ttw^2 \alpha_{WB}-g^2 \alpha_{33},
 \label{tgc} 
\end{eqnarray}
where, 
\begin{eqnarray}
\frac{\delta m_Z^2}{m_Z^2}=    \beta_0 - g^2 \ttw^2\alpha_{WB}+ g^2 \alpha_{33}
\end{eqnarray}
We will show in Appendix~\ref{appA} that  once the  $V^4$ vertices arising from the operators in \eq{oblique}  are also taken into account,  their contribution to the $s^2$ piece in the $V_LV_L \to V_LV_L$ amplitude exactly cancels the TGC contributions  in \eq{tgc}. Thus, the operators in \eq{oblique} ultimately do not give $s^2$ growth  in vector boson scattering  a fact that  can also be understood directly using the goldstone boson equivalence principle. This is, of course, why we do not include these three operators   in our list in Table~\ref{Table1}.

Vertices of the form $V^4$ formed of 4 gauge bosons and no derivatives, are  called anomalous quartic gauge couplings (aQGC). The following list of all possible $U(1)_{em}$ terms of this form was presented in Ref.~\cite{Reuter},
\begin{eqnarray}
\Delta  {\cal L}_{QGC} &=&
   g^2 \ctw^2 \left[ \delta g^{Q}_{ZZ1} Z^\mu Z^\nu W^-_\mu W^+_\nu
    -\delta g^{Q}_{ZZ2} Z^\mu Z_\mu W^{-\nu} W^+_\nu\right]\nonumber\\
    &+&  \frac{g^2}{2}\left[ \delta g ^{Q}_{WW1} W^{-\mu} W^{+\nu} W^-_\mu W^+_\nu
    -\delta g^{Q}_{WW2}\left(W^{-\mu} W^+_\mu\right)^2\right]+\frac{g^2}{4\ctw^4} h^Q_{ZZ} (Z^\mu Z_\mu)^2 \nonumber\\
\end{eqnarray}
We get the following contributions to these from HEFT operators:
\begin{eqnarray}
&& \delta g^{Q}_{ZZ1} =\frac{1}{ \stw^2}\frac{\delta m_Z^2}{m_Z^2}+\frac{g^2}{{4}c_{\theta_w}^4}\left({  4c_2}+c_3\right) , \quad 
	\delta g^{Q}_{ZZ2}=\frac{1}{ \stw^2}\frac{\delta m_Z^2}{m_Z^2}-\frac{g^2}{ 4c_{\theta_w}^4}\left({ 4}c_1+c_4\right), \quad  \nonumber\\
&& \delta g^{Q}_{WW1} =\frac{\ctw^2}{\stw^2}\frac{\delta m_Z^2}{m_Z^2}+g^2c_2 -2 g^2 \alpha_{33}, \quad \delta g^{Q}_{WW2} = \frac{\ctw^2}{\stw^2}\frac{\delta m_Z^2}{m_Z^2}-g^2\left(2c_1+c_2\right)-2 g^2 \alpha_{33}  \nonumber\\&&
	h^Q_{ZZ} = g^2{ \left(4c_1+4c_2+2c_3+2c_4+c_5\right)/4}. 
 \label{qgc} 
\end{eqnarray}

It might seem confusing  at first that while there are only 5 operators  contributing to the $s^2$ piece in the $V_L V_L \to V_L V_L$ forward amplitudes in the HEFT parametrization, there are 8 anomalous couplings that can give such an effect in the $U(1)_{em}$ invariant parametrization.  This, however, does not imply an inconsistency between the two parametrization because only five linear combinations of the 8 anomalous couplings give rise to $s^2$ growth in the forward amplitude. These five linear combinations can be obtained by inverting \eq{tgc} and \eq{qgc} to write the HEFT WCs, $c_1-c_5$, in terms of the above TGCs and aQGCs; this is shown in  Table~\ref{heftmap}. This inversion  also gives us, $\beta_0, \alpha_{WB}$ and $\alpha_{33}$ as a linear combination of anomalous couplings,
\begin{eqnarray}\label{alphas}
  \beta_0&=&  -\ctw^2 \delta g_Z^1 + \left( 1 + \stw^2\right) \delta \kappa^Z -  \stw^2 \delta \kappa^\gamma \nonumber\\
  \alpha_{WB}&=& -\frac{2 \ctw^2}{g^2}\left(\delta g_Z^1 - \delta \kappa^Z + \delta \kappa^\gamma\right)\nonumber\\
  \alpha_{33}&=& \frac{1}{g^2}\left(\ctw^2 \delta g_Z^1 - \ctw^2 \delta \kappa^Z - \stw^2 \delta \kappa^\gamma\right).
\end{eqnarray}
The  right hand sides in \eq{alphas} above are precisely the linear combinations of anomalous couplings that do not give an $s^2$ growth in the forward amplitude.

Next we present the the most general non-redundant list of vertices of the form, $(\partial h)^2V^2$ ,
\begin{eqnarray}
\Delta {\cal L}_{(\partial h)^2V^2}&=&\kappa^{hh}_{WW}\,\frac{h^2}{2v^2}
 {\cal W}^{+\, \mu\nu}{\cal W}^-_{\mu\nu}
+\kappa^{hh}_{ZZ}\,\frac{h^2}{4v^2} Z^{\mu\nu}Z_{\mu\nu}\nonumber\\&+&
\tilde{\kappa}^{hh}_{WW}\,\frac{h^2}{2v^2}
 {\cal W}^{+\, \mu\nu}\tilde{\cal W}^-_{\mu\nu}
+\tilde{\kappa}^{hh}_{ZZ}\,\frac{h^2}{4v^2} Z^{\mu\nu}\tilde{Z}_{\mu\nu}\nonumber\\
&+&
g^{hh}_{Z1} \frac{g^2}{c_{\theta_w}^2} \frac{(\partial_\nu h)^2 Z_\mu Z^\mu}{v^2}  + g^{hh}_{Z2}   \frac{g^2}{c_{\theta_w}^2} \frac{\partial_\mu h \partial_\nu h}{2v^2} Z^\mu Z^{\nu}+g^{hh}_{W1}g^2  \frac{(\partial_\nu h)^2}{v^2} W^{+\mu}W^{-}_{\mu}\nonumber\\
 &+&g^{hh}_{W2} g^2 \frac{\partial_\mu h \partial_\nu h}{2 v^2}  (W^{+\mu}W^{-\nu}+h.c.)
\end{eqnarray}
All other anomalous couplings can be reduced to the above non-redundant set using field-redefinitions and integration by parts. The anomalous couplings  for the contact terms involving field strengths, i.e. $\kappa^{hh}_{WW}$ and $\kappa^{hh}_{ZZ}$, do not contribute to $s^2$ piece of the forward amplitude.\footnote{This can be seen from the fact that the amplitude due to any contact term containing   field strength tensors,  vanishes if we approximate  the longitudinal polarization vector to be $\epsilon^V_{L, \mu}= p_\mu/m_V$. Thus the true  amplitude must be suppressed by powers of $(\epsilon^V_{L, \mu}- p_\mu/m_V)\sim m_V/\sqrt{s}$ and therefore cannot grow as  $s^2$.} The other couplings get   contributions only from the HEFT operators of Table~\ref{Table1} in the unitary gauge,
\begin{eqnarray}
&&g^{hh}_{Z1} = \frac{1}{4}\left(c_{10}+2c_{11} +  c_{12}/2 + c_{13}\right) , \quad 	 g^{hh}_{Z2}=-\frac{1}{2}\left(2c_{10}+c_{12}\right),\nonumber\\
   &&g^{hh}_{W1} =\left(c_{11} +\frac{c_{10}}{2}\right), \quad  ~~~~~~~~~~~~~~~~~~~ g^{hh}_{W2} =-c_{10}, 
 \label{eq:ObsCoeffEWPT} 
\end{eqnarray} 
which clearly  establishes a one to one mapping between these anomalous couplings  and the  HEFT WCs.

As far as couplings of the category $(\partial h)^3V$ are concerned, the most general   lagrangian is, 
\begin{eqnarray}
\Delta {\cal L}^{hV^3} &=& i g \ctw \frac{h}{v} \left[
      {g^{hV^3}_{Z1}} Z_\mu \left(W^+_\nu {\cal W}^{-\mu\nu} - W^-_\nu
         {\cal W}^{+\mu\nu}\right) 
      +  \kappa^{hV^3}_Z W^+_\mu W^-_\nu Z^{\mu\nu}  \right] \nonumber\\
        &+&i g \ctw~\frac{h}{v}\Bigg[\tilde{\kappa}^{hV^3}_Z  W^+_\mu W^-_\nu \tilde{Z}^{\mu\nu}+{\tilde{g}^{hV^3}_{Z1}} Z_\mu W^+_\nu \tilde{{\cal W}}^{-\mu\nu}\Bigg]\nonumber\\
        &+&\frac{h}{v} \left[ g^{hV^3}_{4'}    Z^\mu (W^{+\nu} W^-_{\mu \nu}+h.c.)+g^{hV^3}_5 \epsilon^{\mu \nu \rho \sigma} W^+_{\mu} \lra{D}_\rho W^-_{\nu}Z_\sigma\right] \nonumber\\&+& i {g^{\partial h V^3}_{W1}}\frac{g^3}{2 \ctw}\frac{\partial^\mu h}{v} Z_\nu \left(W^+_\mu W^{-\nu} - h.c.\right)+ {g^{\partial h V^3}_{W2}}\frac{g^3}{2 \ctw}\frac{\partial^\mu h}{v} Z_\nu  (W^+_\mu {  W}^{-\nu}+h.c.)\nonumber\\
       &+&  {g^{\partial h V^3}_{W3}}\frac{g^3}{2 \ctw}\frac{\partial^\mu h}{v} Z_\mu W^+_\mu {W}^{-\mu} +  {g^{\partial h V^3}_{Z}}\frac{g^3}{2 \ctw^3}\frac{\partial^\mu h}{v} Z_\mu Z_\nu Z^{\nu}  
\end{eqnarray}
where we have again removed all possible redundancies. 
 Only the anomalous couplings $g^{\partial h V^3}_{W1},\; g^{\partial h V^3}_{W2},\;g^{\partial h V^3}_{W3}$ and ${g^{\partial h V^3}_{Z}}$ contribute to the $s^2$ piece of the forward amplitude. Once again we obtain a one to one mapping between these anomalous couplings and four of the HEFT operators of Table~\ref{Table1},
\begin{eqnarray}
&&  {g^{\partial h V^3}_{W1}} = \frac{c_8}{4}, \quad
{g^{\partial h V^3}_{W2}} =  \frac{c_7}{2} , \quad 
	 {g^{\partial h V^3}_{W3}} = c_6, \quad  {g^{\partial h V^3}_{Z}}= \frac{2 c_6+2 c_7 + c_9}{4}.  
 \label{eq:ObsCoeffEWPT} 
\end{eqnarray} 

Finally there is one $U(1)_{em}$ invariant operator each in the ${\partial h}^3 V$ and ${\partial h}^4$ categories, 
\begin{eqnarray}
{\cal L}^{h^3V} &=& g^{{\partial h}^3V} \frac{g}{2\ctw v^3}  \partial_\nu h \partial^\nu h \partial_\mu h Z^\mu,~~~~~~~~{\cal L}^{(\partial h)^4} = \frac{g^{(\partial h)^4}}{v^4} \partial_\mu h \partial^\mu h \partial_\nu h \partial^\nu h
\end{eqnarray}
which respectively get   contributions  from $c_{14}$ and $c_{15}$ in Table~\ref{Table1},
\begin{eqnarray}\label{h3v}
g^{{\partial h}^3V} =  -\frac{c_{14}}{2}, \quad    g^{(\partial h)^4} = {c_{15}}.
 \label{eq:ObsCoeffEWPT} 
\end{eqnarray}

\begin{table}[t]
\centering
\renewcommand{\arraystretch}{1.6} 
\begin{tabular}{@{} >{}c<{}  >{}c<{}  @{}}
\hline\hline
\textbf{HEFT Wilson} & \multirow{2}{*}{\textbf{Anomalous Couplings}} \\ 
\textbf{coefficients} & \\ \hline
$c_1$ & $-\frac{1}{2 g^2}\left(\delta g^Q_{WW1} + \delta g^Q_{WW2} - 4 \ctw^2 \delta \kappa^Z - 4 \stw^2 \delta \kappa^\gamma\right)$ \\ 

$c_2$ & $\frac{1}{g^2}\left(\delta g^Q_{WW1} - 2 \ctw^2 \delta \kappa^Z - 2 \stw^2 \delta \kappa^\gamma\right)$ \\ 
$c_3$ & $\frac{4}{g^2}\left(\ctw^4 \delta g^Q_{ZZ1} - \delta g^Q_{WW1} - 2 \ctw^4 \delta g_Z^1 + 2 \ctw^2 \delta \kappa^Z + 2 \stw^2 \delta \kappa^\gamma\right)$ \\ 
$c_4$ & $\frac{2}{g^2}\left(\delta g^{Q}_{WW1} + \delta g^{Q}_{WW2} + \ctw^4\left(4\delta g_{Z}^1-2\delta g^{Q}_{ZZ2}\right)-4\ctw^2\delta \kappa^{Z}-4\stw^2\delta\kappa^{\gamma}\right)$\\
 
$c_5$ & $\frac{2}{g^2}\left(\delta g_{WW1}^{Q} -\delta g^{Q}_{WW2} + 4\ctw^4\left(\delta g^{Q}_{ZZ2} -\delta g^{Q}_{ZZ1}\right) + 2 h^{Q}_{ZZ}\right)$ \\ 
$c_6$ & $g^{\partial h V^3}_{W3}$ \\
$c_7$ & 2$g^{\partial h V^3}_{W2}$ \\ 
$c_8$ & $ 4 g^{\partial h V^3}_{W1}$ \\ 
$c_9$ & $-2\left(2 g^{\partial h V^3}_{W2} + g^{\partial h V^3}_{W3} - 2 g^{\partial h V^3}_{Z} \right)$ \\ 

$c_{10}$ & $- g^{hh}_{W2}$ \\ 

$c_{11}$ & $\frac{1}{2}\left(2  g^{hh}_{W1} + g^{hh}_{W2}\right)$ \\ 

$c_{12}$ & $2\left( g^{hh}_{W2} -  g^{hh}_{Z2}\right)$ \\

$c_{13}$ & $\left(4  g^{hh}_{Z1} +  g^{hh}_{Z2} - 2  g^{hh}_{W1} -  g^{hh}_{W2}\right)$ \\ 

$c_{14}$ & $-2g^{{\partial h}^3V}$ \\ 

$c_{15}$ & $g^{(\partial h)^4}$ \\ \hline\hline
\end{tabular}
\caption{Mapping between the HEFT WCs for the operators in Table~\ref{Table1} and the anomalous couplings presented in Sec.~\ref{anom}.\\
}
\label{heftmap}
\end{table}

In Table~\ref{heftmap} we invert \eq{tgc}-\eq{h3v} to express the HEFT WCs as linear combinations of anomalous couplings presented in this subsection. Our final bounds on $c_1$--$c_{15}$ can be translated to bounds on the anomalous couplings using this table.
\\

\subsection{SMEFT parametrization}
\label{smeftsec}
 A straightforward way to identify  the SMEFT operators contributing to the  gauge-Higgs scattering amplitudes is by  listing   the operators that generate   the anomalous terms presented in the previous subsection.   In the dimension-8 basis  presented in Ref.~\cite{Henning, Li}, the list of operators thus obtained can be subdivided into two groups. First the three dimension 8 operators,  
\begin{eqnarray}
\mathcal{O}_{s1} &=& \left[(D_{\mu}H)^{\dagger}(D_{\nu}H)\right]\left[(D^{\mu}H)^{\dagger}(D^{\nu}H)\right],\\
\mathcal{O}_{s2} &=& \left[(D_{\mu}H)^{\dagger}D^{\mu}H\right]^2,\\
\mathcal{O}_{s3} &=& \left[(D_{\mu}H)^{\dagger}(D_{\nu}H)\right]\left[(D^{\nu}H)^{\dagger}(D^{\mu}H)\right].
\label{smeft3}
\end{eqnarray}
directly generate vertices with four derivatives and four  $h$/Goldstones, thus giving an amplitude that grows as $s^2$ in the forward limit. Here we have defined, $D_\mu H=\partial_\mu +i g \frac{\sigma^a W^a_\mu}{2}H+i g' Y_H B_\mu H$ where $Y_H=1/2$ is the hypercharge of the Higgs doublet.  In addition, the following five operators generating  corrections to vector boson scattering  via input parameter shifts, contributions to the gauge kinetic terms or by directly generating aQGCs, 
\begin{eqnarray}\label{obliquesmeft}
\mathcal{O}_{T} &=& \frac{1}{2} (H^\dagger \lra{D}_\mu H)^2 ,\\
\mathcal{O}_{WB} &=& g g' (H^\dagger \sigma^a H) W^a_{\mu \nu} B^{\mu \nu},\\
\mathcal{O}_{U} &=& g^2 (H^\dagger \sigma^a H) (H^\dagger \sigma^d H) W^a_{\mu \nu} W^{b,\mu \nu},\\
\mathcal{O}_{H^2T} &=&\frac{1}{2}(H^\dagger H) (H^\dagger \lra{D}_\mu H)^2,\\
\mathcal{O}_{H^2 WB} &=& g g'(H^\dagger H)  (H^\dagger \sigma^a H) W^a_{\mu \nu} B^{\mu \nu}.
\end{eqnarray}
 As expected from the goldstone boson equivalence theorem,  these operators do not  give rise to $s^2$ growth in the forward amplitude once we include all the vertices they generate in the unitary gauge. We will show this explicitly in Appendix~\ref{appA}.  

We now obtain the  contributions to the anomalous couplings in Sec.~\ref{anom} from the SMEFT operators listed above. First the TGCs receive the contributions, 
\begin{eqnarray}\label{tgcsmeft}
    && \delta g_{Z}^1 = \frac{1}{2\stw^2}\frac{\delta m_Z^2}{m_Z^2},\quad \delta \kappa^{\gamma} = g^2\left({\cal C}_{WB}+ \frac{{\cal C}_{H^2WB}}{4}\frac{v^2}{\Lambda^2}\right)\frac{v^2}{\Lambda^2}- g^2{\cal C}_U \frac{v^4}{\Lambda^4}, \nonumber \\
    &&\hspace{0.5cm} \delta \kappa^{Z} = \frac{1}{2\stw^2}\frac{\delta m_Z^2}{m_Z^2} -g^2\ttw^2\left({\cal C}_{WB} + \frac{{\cal C}_{H^2WB}}{4}\frac{v^2}{\Lambda^2}\right)\frac{v^2}{\Lambda^2} - g^2{\cal C}_U\frac{v^4}{\Lambda^4}
\end{eqnarray}
where ${\cal C}_i$ is the WC of the SMEFT operator, ${\cal O}_i$, and,
\begin{eqnarray}
\frac{\delta m_Z^2}{m_Z^2} =  -\left({\cal C}_T + \frac{{\cal C}_{H^2T}}{2}\frac{v^2}{\Lambda^2}\right)\frac{v^2}{\Lambda^2}+ 2 g^2t_{\theta_W}^2\left({\cal C}_{WB} + \frac{{ \cal C}_{  H^2WB}}{ 2}\frac{v^2}{\Lambda^2}\right)\frac{v^2}{\Lambda^2} + g^2{\cal C}_U\frac{v^4}{\Lambda^4}\nonumber
\end{eqnarray}
Next the aQGCs get the following contribution from the SMEFT operators, 
\begin{eqnarray}\label{gcsmeft}
&& \delta g^{Q}_{ZZ1} = \frac{1}{s_{\theta_W}^2}\frac{\delta m_Z^2}{m_Z^2} + \frac{g^2}{16\ctw^4}\frac{v^4}{\Lambda^4}({\cal C}_{s1}+{\cal C}_{s3}) , \quad 
	\delta g^{Q}_{ZZ2}=\frac{1}{s_{\theta_W}^2}\frac{\delta m_Z^2}{m_Z^2} - \frac{g^2}{16\ctw^4}\frac{v^4}{\Lambda^4}{\cal C}_{s2}, \nonumber \\
&& \delta g^{Q}_{WW1} = \frac{c_{\theta_W}^2}{s_{\theta_W}^2}\frac{\delta m_Z^2}{m_Z^2} + \frac{g^2}{8}\frac{v^4}{\Lambda^4}{\cal C}_{s1} - 2g^2{\cal C}_U\frac{v^4}{\Lambda^4}, \quad \delta g^{Q}_{WW2} =\frac{c_{\theta_W}^2}{s_{\theta_W}^2}\frac{\delta m_Z^2}{m_Z^2} -\frac{g^2}{8}\frac{v^4}{\Lambda^4}({\cal C}_{s2}+{\cal C}_{s3})- 2g^2{\cal C}_U\frac{v^4}{\Lambda^4}, \nonumber \\
&&	\hspace{5cm}h^Q_{ZZ} = \frac{g^2}{16}\frac{v^4}{\Lambda^4}({\cal C}_{s1}+{\cal C}_{s2}+{\cal C}_{s3}).
\end{eqnarray}
Finally, for the anomalous couplings of the forms, $(\partial h)^2 V^2,~(\partial h)  V^3,~(\partial h)^3 V$  and $(\partial h)^4$, we respectively obtain, 
\begin{eqnarray}\label{hhvvsmeft}
&& \delta g^{hh}_{Z1} = -\frac{1}{8}({\cal C}_{s1}- {\cal C}_{s2} - {\cal C}_{s3})\frac{v^4}{\Lambda^4} , \quad
	\delta g^{hh}_{Z2}=\frac{{\cal C}_{s1}}{2}\frac{v^4}{\Lambda^4}, \nonumber\\
&&\hspace{0.8cm} \delta g^{hh}_{W1} =\frac{{\cal C}_{s2}}{4}\frac{v^4}{\Lambda^4},\quad   \delta g^{hh}_{W2} = \frac{{\cal C}_{s1}+{\cal C}_{s3}}{4}\frac{v^4}{\Lambda^4}, 
\end{eqnarray} 
\begin{eqnarray}
&& { g^{\partial h V^3}_{W1} =\frac{1}{4}\left({\cal C}_{s3} - {\cal C}_{s1}\right)\frac{v^4}{\Lambda^4} },\quad {g^{\partial h V^3}_{W2}} =0 , \quad 
	 {g^{\partial h V^3}_{W3}}=0, \quad  {g^{\partial h V^3}_{Z}}=0,  
 \label{hv3smeft} 
\end{eqnarray} 
and,
\begin{eqnarray}
g^{\partial h^3V} =0, \quad    g^{(\partial h)^4} = \frac{1}{4}\left({\cal C}_{s1} + {\cal C}_{s2} + {\cal C}_{s3}\right)\frac{v^4}{\Lambda^4}.
 \label{hhhsmeft} 
\end{eqnarray} 
  In \eq{gcsmeft}-\eq{hhhsmeft}, the total number of anomalous coupling on the left hand side (18)  is larger than the number of independent SMEFT contributions (6)  on the right hand side.\footnote{Note that the   WCs ${\cal C}_{T},  {{\cal C}_{H^2T}}$ (${\cal C}_{WB},  {{\cal C}_{H^2WB}}$) always appear in these equations in the same linear combination, ${\cal C}_{T} + \frac{{\cal C}_{H^2T}}{2}\frac{v^2}{\Lambda^2}$ (${\cal C}_{WB} + \frac{{\cal C}_{H^2WB}}{  2}\frac{v^2}{\Lambda^2}$). } This implies that the anomalous couplings must satisfy 12 constraints in the dimension-8 SMEFT that can be obtained in a straightforward way by eliminating the SMEFT WCs from \eq{gcsmeft}-\eq{hhhsmeft}. These constraints, expressed  in terms of HEFT WCs, have been shown  in Table~\ref{tab:SMEFT_constraints}.

These constraints imply that in the space of the 15 HEFT WCs of Table~\ref{Table1}, the SMEFT is a three dimensional hyperplane. We will see that the SMEFT positivity bounds   obtained in past literature can  be recovered by  obtaining a projection of our  15 dimensional bounds  derived in Sec.  on the 3 dimensional SMEFT plane using Table~\ref{tab:SMEFT_constraints}.
\begin{table}[t]
\centering
\renewcommand{\arraystretch}{1.4} 
\begin{tabular}{@{}>{}c<{} @{}} 
\hline \hline
\text{SMEFT Constraints} \\ \hline
$c_3 + c_4 = 0$ \\ 
 $c_5 = 0$ \\ 
$16c_1  - 2c_{10} - 4c_{11} - c_{12} - 2c_{13} = 0$ \\ 
$c_{10} + c_{12}/2 + 4c_2 = 0$\\
$4c_1 - c_{10}/2 -c_{11} + c_4 = 0$ \\
$c_{10} + 4c_2 + c_3 = 0$ \\ 
$4c_3 - c_8 = 0 $ \\
$c_6,\; c_7, \; c_9=0$ \\ 
$c_{14} = 0$ \\ 
$4(c_1+c_2) - c_{15} = 0$\\
\hline\hline
\end{tabular}
\caption{Linear constraints on the WCs of NLO  HEFT operators implied by the SMEFT  truncated at dimension-8 level.\label{tab:SMEFT_constraints}}
\end{table}

Finally we mention a few subtleties before proceeding further. First of all the constraints obtained in Table~\ref{tab:SMEFT_constraints} will be broken by higher dimension   operators suppressed by additional powers of the cutoff.  We assume the following power-counting for  the SMEFT WCs, 
\begin{eqnarray}\label{smeftpc}
    {\rm SMEFT}&:&  \frac{\Lambda^4}{g_H^2}  \left(\frac{D}{\Lambda}\right)^{n_D} \left(\frac{g_H H}{\Lambda}\right)^{n_H}
\end{eqnarray}
where, $g_H < \Lambda/v$, is the coupling of the Higgs to heavy BSM states. The   contribution of  higher dimensional WCs to the scattering amplitudes relative to that of dimensional-8 ones is therefore suppressed by powers of  $g_H^2 v^2/\Lambda^2$. We see that in the limit $g_H \to \Lambda/ v$ this suppression vanishes and indeed the SMEFT power counting in \eq{smeftpc} approaches the HEFT power counting in \eq{pc}. This is, however, exactly the limit in which the SMEFT expansion breaks down and the correct effective description is given by HEFT.   A second subtlety is that if we assume the power counting scheme in \eq{smeftpc},  the contribution of diagrams with two insertions of dimension-6 operators cannot be ignored for a $g_H$ that is  ${\cal O}(1)$. Fortunately, however, the dimension-6 contributions  have been found to give a negative definite contribution~\cite{zhang1,zhang2} to the $s^2$ piece of the forward amplitude so that individual bounds can still be placed on the dimension-8 contribution. Thus in our list above we have only included  SMEFT operators that contribute to gauge-Higgs scattering with a single insertion.


\subsection{Longitudinal gauge-Higgs scattering amplitude}
\label{amp}
We now present the  amplitude for longitudinal gauge-Higgs scattering, $b_i b_j \to b_k b_l$ with the indices taking values from 1--4 and $b=\{W_{1L}, W_{2L}, Z_{L}, h\}$. We will show using only $U(1)_{em}$ invariance that the coefficient of $s^2$ in the forward amplitude can be parametrized by 15 independent parameters. We will then provide an explicit mapping between these parameters and the 15 HEFT WCs---or equivalently the 15 linear combinations of anomalous couplings in Table~\ref{heftmap}.

We separate the full gauge-Higgs scattering  amplitude into two parts,
\begin{eqnarray}
\mathcal{M}^V_{ijkl}(s,t)=\mathcal{M}^{V,sing}_{ijkl}(s,t)+\mathcal{\tilde{M}}^V_{ijkl}(s,t)
\label{amp2term}
\end{eqnarray}
where  the first term contains all the IR singularities from from tree-level exchange of light particles in the $s,t$ or $u$ channel as well as loop contributions. For the goldstone scattering amplitude, $\phi_i \phi_j \to \phi_k \phi_l$,  we can similarly subtract out the IR singularities to obtain $\mathcal{\tilde{M}}^\phi_{ijkl}$. The goldstone boson equivalence theorem implies, $\mathcal{\tilde{M}}^V_{ijkl}=\mathcal{\tilde{M}}^\phi_{ijkl}$ for  $s \gg m_{h,W,Z}^2$ which allows us to drop the superscripts and simply use $\mathcal{\tilde{M}}_{ijkl}$ to denote the amplitude. In the UV, $s > \Lambda^2$, $\mathcal{\tilde{M}}_{ijkl}$  gives the exact amplitude. For $m_{h, W, Z}^2\ll  s, t \ll \Lambda^2$ we can expand the second term in powers of $s$ and $t$ as follows, 
\begin{eqnarray}\label{eftst}
    \mathcal{\tilde{M}}_{ijkl}(s,t)= \sum_{m,n} c^{m,n}_{ijkl} s^m t^n
\end{eqnarray}
where we have assumed that the tree-level HEFT contribution is a good approximation for the low-energy amplitude.\footnote{See Ref.~\cite{rattazzi, Riembau, Chala} for a discussion about the effect of EFT-loops on positivity bounds. }  For the purpose of deriving the positivity bounds in this work we require only,  $c^{2,0}_{ijkl}$. 

Before producing a mapping between the, $c^{2,0}_{ijkl}$, and the HEFT WCs let us first count the number of independent, $c^{2,0}_{ijkl}$,  imposing only $U(1)_{em}$ invariance. As the $U(1)_{em}$ acts like an  $SO(2)$ on the indices, $i=1,2$, we write  the amplitude, $\mathcal{M}_{ijkl}$, using invariant tensors of the $SO(2)$ symmetry group,
\begin{eqnarray}
    \mathcal{\tilde{M}}_{ijkl} = \delta_{ij}\delta_{kl}~f(s,u) + \delta_{ik}\delta_{jl}~f(t,u) + \delta_{il}\delta_{kj}~f(u,s)\hspace{0.5cm} {\rm for~} i,j,k,l\in[1,2],\label{symform}
\end{eqnarray}
where, the function $f(s,u)$ is symmetric under exchange of $s$ and $u$.  The   form of the amplitude  in \eq{symform} implies the following  constraints,
\begin{eqnarray}
\mathcal{\tilde{M}}_{1111}(s) &=&\mathcal{\tilde{M}}_{2222}(s) \label{1stcons} \\
    \mathcal{\tilde{M}}_{1111}(s) &=& \mathcal{\tilde{M}}_{1212}(s)+\mathcal{\tilde{M}}_{1221} + \mathcal{\tilde{M}}_{1122}(s). \label{2ndcons}
\end{eqnarray}
For  2-to-2 scattering amplitudes involving $Z_L$  and $h$ (or $\phi_3$ and $h$ for the corresponding goldstone amplitude),  $U(1)_{em}$ invariance implies $\mathcal{\tilde{M}}^V_{12ii}=\mathcal{\tilde{M}}^V_{1ii2}=\mathcal{\tilde{M}}^V_{1i2i}=0$, $\mathcal{\tilde{M}}^V_{1i1i} = \mathcal{\tilde{M}}^V_{2i2i}$,  $\mathcal{\tilde{M}}^V_{11ii} = \mathcal{\tilde{M}}^V_{22ii}$ and $\mathcal{\tilde{M}}^V_{1ii1} = \mathcal{\tilde{M}}^V_{2ii2}$ where $i=3,4$.

These constraints can be directly translated to the, $c^{2,0}_{ijkl}$,  given by,
\begin{equation}
   c^{2,0}_{ijkl}= \left. \frac{1}{2}\frac{\partial^2 \tilde{\cal M}_{ijkl}}{\partial s^2}\right \vert_{s,t=0}.
\end{equation}
They would imply that many of the  $c^{2,0}_{ijkl}$ are equal to each other and many others vanish. In matrix form we can write, 
\begin{equation}\label{c20}
v^4 c^{2,0}_{ijkl}=   \left(
\begin{array}{cccccccccccccccc}
 a_1 & 0 & 0 & 0 & a_2 & 0 & 0 & 0 & a_3 & 0 & 0 & a_9 & 0 & 0 & a_9 & a_{12} \\
 0 & a_4 & 0 & a_2 & 0 & 0 & 0 & 0 & 0 & 0 & 0 & -a_{16} & 0 & 0 & a_{16} & 0 \\
 0 & 0 & a_6 & 0 & 0 & 0 & a_3 & 0 & 0 & a_{10} & 0 & 0 & a_9 & a_{16} & 0 & 0 \\
 0 & a_2 & 0 & a_4 & 0 & 0 & 0 & 0 & 0 & 0 & 0 & a_{16} & 0 & 0 & -a_{16} & 0 \\
 a_2 & 0 & 0 & 0 & a_1 & 0 & 0 & 0 & a_3 & 0 & 0 & a_9 & 0 & 0 & a_9 & a_{12} \\
 0 & 0 & 0 & 0 & 0 & a_6 & 0 & a_3 & 0 & 0 & a_{10} & 0 & -a_{16} & a_9 & 0 & 0 \\
 0 & 0 & a_3 & 0 & 0 & 0 & a_6 & 0 & 0 & a_9 & a_{16} & 0 & a_{10} & 0 & 0 & 0 \\
 0 & 0 & 0 & 0 & 0 & a_3 & 0 & a_6 & 0 & -a_{16} & a_9 & 0 & 0 & a_{10} & 0 & 0 \\
 a_3 & 0 & 0 & 0 & a_3 & 0 & 0 & 0 & a_5 & 0 & 0 & a_7 & 0 & 0 & a_7 & a_8 \\
 0 & 0 & a_{10} & 0 & 0 & 0 & a_9 & -a_{16} & 0 & a_{14} & 0 & 0 & a_{12} & 0 & 0 & 0 \\
 0 & 0 & 0 & 0 & 0 & a_{10} & a_{16} & a_9 & 0 & 0 & a_{14} & 0 & 0 & a_{12} & 0 & 0 \\
 a_9 & -a_{16} & 0 & a_{16} & a_9 & 0 & 0 & 0 & a_7 & 0 & 0 & a_{13} & 0 & 0 & a_{15} & a_{11} \\
 0 & 0 & a_9 & 0 & 0 & -a_{16} & a_{10} & 0 & 0 & a_{12} & 0 & 0 & a_{14} & 0 & 0 & 0 \\
 0 & 0 & a_{16} & 0 & 0 & a_9 & 0 & a_{10} & 0 & 0 & a_{12} & 0 & 0 & a_{14} & 0 & 0 \\
 a_9 & a_{16} & 0 & -a_{16} & a_9 & 0 & 0 & 0 & a_7 & 0 & 0 & a_{15} & 0 & 0 & a_{13} & a_{11} \\
 a_{12} & 0 & 0 & 0 & a_{12} & 0 & 0 & 0 & a_8 & 0 & 0 & a_{11} & 0 & 0 & a_{11} & a_{15} \\
\end{array}
\right)\nonumber\\
\end{equation}
whre the rows and columns are ordered as follows,
\begin{equation}
\{11,12,13,21,22,23,31,32,33, 14,2 4, 3 4,4 1, 4 2, 4 3, 44\}
\end{equation}
 and respectively denote different possible initial and final states for gauge-Higgs scattering. For instance, the top left entry $a_1$ corresponds to either the    $W_{1L} W_{1L} \to W_{1L} W_{1L}$. Our parametrization in terms of the $a_1-a_{16}$ already incorporates all the constraints arising from $U(1)_{em}$ apart from \eq{2ndcons} which implies the relation, 
 \begin{eqnarray}\label{aconst}
   a_1=2 a_2 +a_4.  
 \end{eqnarray}
There are thus 15 independent parameters required to  completely parametrize the $s^2$ piece of the forward amplitude.\footnote{Note that the coefficients of $st$ and $t^2$ in $\tilde{\cal M}_{ijkl}$ are  related to the $c^{2,0}_{ijkl}$ by crossing-symmetry and are thus not independent. Thus we need only 15 parameters to parametrize all the $c^{m,n}_{ijkl}$ with $m+n=2$.}

We now proceed to calculate the amplitude, $\mathcal{\tilde{M}}_{ijkl}$,  using the lagrangians presented in the previous subsections. We can either use the unitary gauge parametrization of Sec.~\ref{anom} or  the HEFT parametrization in Sec.~\ref{HEFT_param} to obtain the  matrix  in \eq{c20}.   We carry out both these computations and explicitly verify that  they yield the same answer. We find,
\begin{eqnarray}\label{amap}
&& a_1=16 \left(c_1+c_2\right),\quad
a_2=4 \left(2 c_1+c_2\right), \quad
a_3=8 c_1+4 c_2+c_3+2 c_4\nonumber\\
&&a_4=8 c_2,\quad
a_5=16 \left(c_1+c_2\right)+8 \left(c_3+c_4\right)+4 c_5,\quad
a_6=2 \left(4 c_2+c_3\right)\nonumber\\
&&a_7=2 c_6+2 c_7+c_9,\quad
a_8=2 c_{11}+c_{13},\quad
a_9=c_6+\frac{c_7}{2},\quad
a_{10}= c_7\nonumber\\
&&a_{11}= -\frac{c_{14}}{2},\quad
a_{12}= 2 c_{11},\quad
a_{13}= -2 c_{10} -c_{12},\quad
a_{14}= -2c_{10},\nonumber\\
&&\hspace{3.5cm}a_{15}=  4 c_{15},\quad
a_{16}= \frac{c_8}{4}.
\end{eqnarray}
 We see that \eq{amap} clearly satisfies \eq{aconst} and thus provides a one-to-one mapping between the 15 independent parameters of the amplitude and the 15 HEFT WCs---and hence the 15 linear combination of  the anomalous couplings in Table~\ref{heftmap}


Finally note that from \eq{amap} it is clear that the $c^{2,0}_{ijkl}$ get  no contribution from the HEFT WCs in \eq{oblique} (and thus no contribution from  the SMEFT WCs in \eq{obliquesmeft}). While this fact is  trivial for goldstone scattering, $\phi_i \phi_j \to \phi_k \phi_l$ it is, far from   obvious in the unitary gauge calculation involving anomalous couplings. We, therefore, carry out the latter  calculation explicitly in Appendix~\ref{appA}.
 
\section{The positivity cone: analytical positivity constraints}\label{three}


In this section we derive analytical positivity bounds on the operators in Table~\ref{Table1}. Our starting point would be the twice-subtracted dispersion relation for longitudinal gauge-Higgs scattering at a fixed $t$. In order to derive this we assume that for a given $t$,  $\tilde{\mathcal{M}}_{ij}(s,t)$ defined in \eq{amp2term} is an analytic function in the complex $s$-plane apart from $s$ and $u$-channel singularities on the real line  due to new states at $s\geq \Lambda^2$ and  $u\geq \Lambda^2$. We  also  make use  of the  Froissart-Martin bound~\cite{Froissart, Martin} which not only restricts the high energy behavior of the full amplitude but can also be  applied directly to  $\tilde{\mathcal{M}}_{ij}(s,t)$.  This is because of  the fact that in the  HEFT at NLO, the residue of   $t$-channel poles below the cutoff has at most a linear dependence on $s$.\footnote{The photon exchange diagrams may still seem problematic as they   blow up  in the $t \to 0$ limit. One can, however, circumvent this issue because positivity arguments also hold away from the forward limit for a   small but finite $t$.}
Finally, we  use  $\mathcal{\tilde{M}}_{ijkl}(s+i \epsilon, t)=\mathcal{\tilde{M}}^*_{ijkl}(m_{ijkl}^2-s-i \epsilon-t, t)$  which follows from  $su$-crossing symmetry and real-analiticity where $m_{ijkl}^2 = m_i^2 + m_j^2 + m_k^2 + m_l^2$.  We can then use the standard procedure to obtain the dispersion relation (see for eg. Ref.~\cite{deRham}), 
\begin{eqnarray}\label{disp}
 \mathcal{\tilde{M}}_{ijkl}&=& a^{(0)}_{ijkl}(t) +a^{(1)}_{ijkl}(t) s
 \nonumber\\&+&\frac{(s-s_\star)^2}{2\pi i} \int\limits_{\Lambda^2}^{\infty}ds^{\prime}\frac{1}{(s'-s_\star)^2}\left(\frac{{\rm Disc~}\mathcal{\tilde{M}}_{ijkl}(s^{\prime},t)}{(s^{\prime}-s)}+\frac{{\rm Disc~}\mathcal{\tilde{M}}_{ilkj}(s^{\prime},t)}{(s^{\prime}-u)}\right)\label{eq2.5}
\end{eqnarray}
where ${\rm Disc~} {M}_{ijkl} = \lim\limits_{\epsilon\to 0^{+}}\mathcal{M}_{ijkl}(s+i\epsilon,t) - \mathcal{M}_{ijkl}(s-i\epsilon,t)$ and $s_*$  is an arbitrary subtraction point   that we will choose to be $s_*=t/2-m_{ijkl}^2$.

To derive  positivity constraints of the kind derived in Ref.~\cite{Allan_Adams_2006}, it is enough to consider the  $ij \to ij$ process in the forward limit, $t \to 0$.  We can then  use the optical theorem, ${\rm Disc~} {M}_{ijij} = 2i~{\rm Im} {M}_{ijij}= \sqrt{(s^{\prime}-(m^{+}_{ij})^2)(s^{\prime}-(m^{-}_{ij})^2)}\sigma_{ij}(s^{\prime})$ and  \eq{disp} to obtain for the forward amplitude $\tilde{\mathcal{M}}_{ij}(s)=\tilde{\mathcal{M}}_{ijij}(s,t\to 0)$,
\begin{eqnarray}
\frac{1}{2}\frac{\partial^2}{\partial s^2}\tilde{\mathcal{M}}_{ij}(s) &=& \frac{1}{\pi}\int\limits_{\Lambda^2}^{\infty}ds^{\prime}\sqrt{(s^{\prime}-(m^{+}_{ij})^2)(s^{\prime}-(m^{-}_{ij})^2)}\sigma_{ij}(s^{\prime})\left[\frac{1}{(s^{\prime}-s)^3}\right.\\
&&\hspace{2cm}\left.+\frac{1}{(s^{\prime}-(m^{+}_{ij})^2+s)^3}\right]\label{positivity}
\end{eqnarray}
where the right hand side  is clearly positive. Here, we have defined $m^{\pm}_{ij} = m_i \pm m_j$. We can immediately  use the above equation to obtain our first set of positivity bounds by demanding that the diagonal elements of the matrix in \eq{c20} are positive.

More optimal bounds can  be obtained, by considering the  scattering of the quantum superposed states, $|\alpha\rangle=\alpha_i |i\rangle, |\beta\rangle=\beta_j |j\rangle$,
\begin{equation}
|\alpha; k_1\rangle + |\beta; k_2\rangle \to |\alpha; k_3\rangle + |\beta; k_4\rangle.
\end{equation}
which has the following amplitude,
\begin{equation}
\tilde{\mathcal{M}}_{\bm{\alpha}\bm{\beta}\to\bm{\alpha}\bm{\beta}}(s,t)=\sum_{i,j,k,l=1}^3\alpha_{i}\beta_{j}\alpha^{\ast}_{k}\beta^{\ast}_{l}\tilde{\mathcal{M}}_{ijkl}(s,t)\label{amp9}
\end{equation}
Using the arguments that led to \eq{positivity} we now infer, 
\begin{equation}
\frac{\partial^2}{\partial s^2}\tilde{\mathcal{M}}_{\bm{\alpha}\bm{\beta}\to\bm{\alpha}\bm{\beta}}(s,t\to 0)\vert_{s=0}=\sum_{i,j,k,l=1}^3\alpha_{i}\beta_{j}\alpha^{\ast}_{k}\beta^{\ast}_{l}\frac{\partial^2}{\partial s^2}\tilde{\mathcal{M}}_{ijkl}(s,t\to 0)\vert_{s=0}>0 \label{amp10}
\end{equation}
This implies that the matrix,
\begin{eqnarray}\label{gamma}
   \left(\gamma_{\bm{\beta}}\right)_{ik}= \beta_{j}\beta^{\ast}_{l}\frac{\partial^2}{\partial s^2}\tilde{\mathcal{M}}_{ijkl}(s,t\to 0)\vert_{s=0}\label{pos_cond}
\end{eqnarray}
is a positive definite hermitian matrix which in turn requires that all its  principal minors must be positive.  In particular, we can demand the positivity of all the  diagonal entries of $\left(\gamma_{\bm{\beta}}\right)_{ik}$ as well as the determinant of all principal  2$\times$ 2 sub-matrices. This yields  the 10 conditions presented in  Appendix~\ref{beta} that  depend on the $\beta_i$ in \eq{gamma}. In  Appendix~\ref{beta} we show how the  $\beta_i$ can be optimally chosen to obtain the 16 positivity constraints presented in Table~\ref{relations}. 

In the upper block of Table~\ref{relations} we present the first set of positivity constraints obtained from requiring the positivity of the diagonal elements of $\gamma_{\bm{\beta}}$. These constraints involve only the CP even WCs, $c_{1-5}, c_{10}, c_{12}$, and $c_{15}$, that  contribute to the elastic scattering processes,  $V_Lh \to V_Lh,~Z_LZ_L \to Z_LZ_L, W_LW_L \to W_LW_L, W_LZ_L \to W_LZ_L$ and $hh\to hh$.  Positivity bounds on the CP-odd WCs or  even the CP-even ones contributing to the inelastic processes, $V_LV_L, hh\to V_Lh$, arise from the second set of constraints obtained by requiring the positivity of the determinants of the 2$\times$2 principal sub-matrices; these have been presented in the lower block of Table~\ref{relations}.  The second set of constraints has the form, $A^2 < BC$, where $B$ and $C$ are linear combinations of CP-even WCs contributing  to elastic processes---that are already required to be positive by the first set of positivity constraints---and $A$ is a linear combination that  includes $CP$-odd couplings and WCs contributing to $V_LV_L, hh\to V_Lh$. The second set of conditions thus imply that all $CP$-odd couplings  and all the WCs contributing to $V_LV_L, hh\to V_Lh$ must vanish if the other WCs  vanish. This also implies that  theoretical or experimental bounds  on the CP even couplings contributing to the elastic processes imply upper  bounds on the magnitude of the $CP$-odd couplings  and the WCs contributing to the inelastic processes, $V_LV_L, hh\to V_Lh$---a fact that we will use to compute our final positivity constraints in Sec.~\ref{hefthedron}.

\begin{table}[t]
\centering
\renewcommand{\arraystretch}{1.2} 
\begin{tabular}{@{} >{}c<{}@{}}
\hline \hline
\textbf{Set 1} \\ \hline
    $c_2 > 0$ \\  
    $c_1 + c_2 > 0$ \\ 
    $4(c_1 + c_2) + 2(c_3 + c_4) + c_5 > 0$ \\ 
    $4c_2 + c_3 > 0$ \\ 
    $c_{10} < 0$ \\  
    $2c_{10} + c_{12} < 0$ \\ 
    $c_{15} > 0$\\ \hline 
\textbf{Set 2}\\ \hline
    $4a_3^2 < (a_6 + \sqrt{a_1a_5})^2$ \\ 

    $a_9^2 < a_{6}a_{11}$\\
    
    $a_{12}^2 < a_{5}a_{13}$\\
    
    $a_{15}^2 < a_{13}a_{16}$ \\    
    
    $4a_7^2 < \left(\sqrt{a_{11}a_6} + \sqrt{a_{13}a_1}\right)^2$ \\ 
    
    $4a_{10}^2 < \left(\sqrt{a_6a_{11}} + \sqrt{a_4a_{13}}\right)^2$ \\  
    $4a_8^2 < \left( a_{11} + \sqrt{a_1a_{16}}\right)^2$ \\  
    $\left(a_{12} + a_{15} + 2a_{14}\right)^2 < (a_5+a_{13}+2a_{12})(a_{13}+a_{16}+2a_{15})$\\ \hline\hline
\end{tabular}
\caption{Positivity constraints on the HEFT WCs in Table~\ref{Table1}. These constraints can be translated to constraints on the anomalous couplings contributing to longitudinal gauge-Higgs scattering using the mapping presented in Table~\ref{heftmap}. Here $a_1=16(c_1+c_2),~ a_2=4(2c_1+c_2),~ a_3=8c_1 + 2c_2 + c_3 + 4c_4,~
a_4=8c_2, ~ a_5=16(c_1+c_2)+8(c_3+c_4)+4c_5,~ a_6=2(4c_2+c_3),~
a_7=c_6+\frac{c_7}{2},~ a_8=2c_{11},~ a_9=c_7, ~ a_{10}=-\frac{c_8}{4},~ a_{11}=-2c_{10},~
a_{12}=2(c_6+c_7+c_9/2),~ a_{13}=-(2c_{10}+c_{12}),~ a_{14}=(2c_{11}+c_{13}),~
a_{15}=-c_{14}/2,~ a_{16}=4c_{15}.$}
   \label{relations}
\end{table}


Let us now compare bounds obtained in Table~\ref{relations} with existing literature. Some positivity constraints on the electroweak chiral Lagrangian were  earlier explored in  Ref.~\cite{ewcl1, ewcl2}. The  first two inequalities presented in Table~\ref{relations} are consistent with the results of these studies. Furthermore, positivity constraints on the SMEFT lagrangian have been derived in Ref.~\cite{zhang1, zhang2, remmen1}.  We can recover the SMEFT positivity constraints by expressing the HEFT WCs in terms of the SMEFT ones as follows, 
\begin{eqnarray}
 c_1 &=& \frac{1}{16}({\cal C}_{s3} + {\cal C}_{s2} - {\cal C}_{s1})\frac{v^4}{\Lambda^4}, \quad  
 c_2 =  \frac{{\cal C}_{s1}}{8} \frac{v^4}{\Lambda^4}, \quad  
 c_3 =  \frac{{\cal C}_{s3} - {\cal C}_{s1}}{4} \frac{v^4}{\Lambda^4}=-c_4, \quad
c_5 = 0, \nonumber
\end{eqnarray}
where  \(\Lambda\) denotes the cutoff scale. Using the above equation and Table~\ref{relations}, we can derive the following constraints on the SMEFT parameter space:
\begin{eqnarray}\label{smeftcone}
 {\cal C}_{s1} > 0, \quad  
 {\cal C}_{s1} + {\cal C}_{s3} > 0, \quad  
 {\cal C}_{s1}+ {\cal C}_{s2} + {\cal C}_{s3}> 0.
\end{eqnarray}
This reproduces the bounds reported in \cite{zhang2}. As  shown in Ref.~\cite{zhang2}, contributions to the  scattering amplitudes from two dimension-6 insertions do not invalidate the above bounds as they  turn out to be negative definite.   

 Together, the constraints in Table~\ref{relations} give an allowed region that forms a convex cone~\cite{cone} in the space of WCs.  Note that the positive definiteness of $\gamma_{\bm{\beta}}$ implies further constraints in  addition to those in Table~\ref{relations} as we can further demand, 
\begin{eqnarray}
\det \left(\gamma_{\bm{3\beta}}\right)>0\nonumber\\
\det \left(\gamma_{\bm{\beta}}\right)>0
\label{further}
\end{eqnarray}
where $\gamma_{\bm{3\beta}}$ is the   submatrix of $\gamma_{\bm{\beta}}$ obtained by removing the fourth row and column. These conditions must hold for an arbitrary choice of the  $\beta_{i}$. It is, however, not straightforward to optimally choose the $\beta_i$ to get  analytical bounds. We will not explore these two conditions further as the numerical  methods discussed in the next section provide an alternative way to derive the positivity cone that includes the implications of both Table~\ref{relations} and  \eq{further}. Furthermore, they also provide double-sided bounds  on the WCs that close this conical allowed region. In the coming sections we will refer to the region in the HEFT space consistent with Table~\ref{relations} as the `HEFT positivity cone' and the region defined by \eq{smeftcone} as the `SMEFT positivity cone'.

\section{Capping the positivity cone: double sided bounds from $s$-$t$ crossing and unitarity.}
\label{dsided}

In the previous section we  derived the positivity bounds summarized in Table~\ref{relations} and \eq{further}.  In  order to derive these constraints, we used \( su \) crossing symmetry but did not impose \( st \) crossing symmetry. As we will discuss shortly, the full implications of unitarity were also not applied in Sec.~\ref{positivity}. These additional conditions  can be imposed using the numerical methods developed in \cite{Extremal, Tolley, Chen:2023bhu, Hong:2024fbl}, that result in  double-sided bounds on EFT WCs. We will now show how these methods can be used to close the conical allowed region obtained in  the previous section.  To use the terminology introduced in Ref.~\cite{Chen:2023bhu, Hong:2024fbl}  these numerical methods alow us to `cap the positivity cone'(from here on we will sometimes refer to  these numerical bounds  as `capping bounds'. We will carry out this procedure   for the WCs contributing to the $V_LV_L\to V_LV_L,\;hh$ and $hh\to hh$ processes. Below, we briefly outline the procedure; for further details, please see \cite{Chen:2023bhu} and references therein.

\paragraph{IR-UV relations} First, let us recast \eq{disp} in terms of the new variable,  $v=s+t/2 +m_{ijkl}^2/2=-u-t/2- m_{ijkl}^2/2$,
\begin{eqnarray}
 \mathcal{\tilde{M}}_{ijkl}&=& \tilde{a}^{(0)}_{ijkl}(t) +\tilde{a}^{(1)}_{ijkl}(t) v
 +\frac{(s-s_\star)^2}{2\pi i} \int\limits_{\Lambda^2}^{\infty}ds^{\prime}\frac{1}{(s'-s_\star)^2}\left(\frac{{\rm Disc}\mathcal{\tilde{M}}_{ijkl}(s^{\prime},t)}{(s^{\prime}-s)}+\frac{{\rm Disc}\mathcal{\tilde{M}}_{ilkj}(s^{\prime},t)}{(s^{\prime}-u)}\right)\nonumber\\\label{eq4.1}
\end{eqnarray}
 and also rewrite the low-energy expansion for the EFT amplitude in terms of $t$ and the new variable $v$,   
\begin{equation}\label{eftvt}
    \tilde{\mathcal{M}}_{ijkl}(s,t)  =   \sum_{m,n} \tilde{c}_{ijkl}^{m,n} v^m t^n.
\end{equation}
Taking $m\geq 2$ derivatives with respect to $v$ and $n$    derivatives with respect to $t$, we obtain, 
\begin{eqnarray}
    \tilde{c}_{ijkl}^{m,n} & =& \Bigg\langle  \big[\rho_l^{ijkl}(s^{\prime}) + (-1)^m \rho_l^{ilkj}(s^{\prime}) \big]\sum_{p=0}^{n}\frac{L_l^p H_{m+1}^{n-p}}{{s^{\prime}}^{m+n+1}} \Bigg\rangle\
     \label{iruv}
    \end{eqnarray}
where  $\rho^{ijkl}_l$ are the `spectral densities' obtained from the partial wave expansion of the  absorptive part of the high energy amplitude, 
\begin{eqnarray}\label{spec}
\frac{1}{2i}\mathrm{Disc}\, \tilde{\cal M}_{ijkl}(\mu,\,t)=16\pi \sum_{l=0}^\infty(2l+1)P_l\left(1+\frac{2t}{\mu}\right)\rho_l^{ijkl}(\mu)\,.
\end{eqnarray}

$P_l$ are  the Legendre polynomials and we have used the   notation of Ref.~\cite{Chen:2023bhu}, where, 
\begin{eqnarray}
 \Big\langle\,...\,\Big\rangle &=& \sum_l 16(2l+1) \int_{\Lambda^2}^{\infty} ds^{\prime}\Big(\,...\,\Big)
 \end{eqnarray}
 and,
 \begin{eqnarray}
 L_l^n &=& \frac{\Gamma(l+n+1)}{n!\Gamma(l-n+1)\Gamma(n+1)},
H_{m+1}^q =\frac{\Gamma(m+q+1)}{(-2)^q\Gamma(q+1)\Gamma(m+1)}.
\end{eqnarray}

Eq.~(\ref{iruv}) relates the  the low-energy EFT amplitude to the spectral densities, $\rho^{ijkl}_l$, which encode UV dynamics. Note that   with our assumption that EFT loops can be neglected,  the HEFT WCs can be written as linear combinations of the $\tilde{c}^{mn}_{ijkl}$. We can thus  find their  allowed range by varying the spectral densities in the range allowed by unitarity. Apart from   unitarity constraints, the spectral densities  must also satisfy the so-called null-constraints that arise from $st$-crossing,  and symmetry constraints  implied by the unbroken $U(1)_{em}$.  Note that the positivity constraints derived in the last section arise from the fact that for certain linear combinations of the $\tilde{c}_{ijkl}^{m,n}$ in \eq{iruv}, the right hand side can be related to a cross-section by the optical theorem. Therefore, the analytical positivity constraints of Table~\ref{relations} would be automatically satisfied by the allowed region obtained using the numerical procedure described in this section.

\paragraph{Null constraints}
 We now derive the consequences of  imposing $st$ crossing symmetry on the space of WCs and thus on the spectral densities. For the EFT amplitude in \eq{eftst}, $st$ crossing would imply,
 \begin{eqnarray}
   c^{mn}_{ijkl} = c^{nm}_{ikjl}.
 \end{eqnarray}
By relating the expansions in \eq{eftst} and \eq{eftvt} we can rewrite the above condition in terms of the $\tilde{c}^{m,n}_{ijkl}$ to obtain, 
\begin{align}
    \mathcal{N}_{ijkl}^{m,n}
&=\sum_{a=m}^{m+n}\frac{\Gamma(a+1)\tilde{c}_{ijkl}^{a,m+n-a}}{2^{a-m}\Gamma(m+1)\Gamma(a-m+1)} - \sum_{a=n}^{m+n}\frac{\Gamma(a+1)\tilde{c}_{ikjl}^{a,m+n-a}}{2^{a-n}\Gamma(n+1)\Gamma(a-n+1)}=0
  \label{ncset1}
\end{align}
Now one can substitute $\tilde{c}_{ijkl}^{m,n}$s in \eq{ncset1} in terms of $\rho_l^{ijkl}$s using \eq{iruv}. to obtain the first set of null constraints on the spectral densities~\cite{Tolley, Extremal}. A second set of null constraints can be derived by noting that $su$-crossing symmetry implies 
\begin{eqnarray}
 {\tilde{c}_{ijkl}^{1,n}+\tilde{c}_{ilkj}^{1,n}}=0.
\label{ncset12}  
\end{eqnarray}
Although  $\tilde{c}^{1,n}$  cannot be directly connected to the spectral densities via \eq{eq4.1},  the first set of null constraints allows us to write them in terms of  $\tilde{c}^{m\geq2,n'}$s  and thus can be expressed in terms of spectral densities.

\paragraph{Unitarity constraints} In addition to the null-constraints, the spectral densities must obey   unitarity constraints. The set of unitarity constraints that have been utilized in this work are the following,
\begin{eqnarray}
    &&0\leq \rho_l^{iiii}(s) \leq2\\
    &&0\leq \rho_l^{ijij}(s) \leq\frac{1}{2}\\
    &&-1\leq \rho_l^{iijj}(s)\leq 1\\
    &&\vert \rho_i^{iijj}(s) \vert \leq 1 - \left\vert1-\frac{\rho_l^{iiii}(s) + \rho_l^{jjjj}(s)}{2}\right\vert.
\end{eqnarray}\label{Unitarity_cons}
The above inequalities were derived and  discussed in great detail in Ref.~\cite{Chen:2023bhu}. 

\paragraph{Symmetry constraints} Apart from crossing symmetry and unitarity,    we  can impose  additional symmetry constraints due to the unbroken $U(1)_{em}$  group.  Using Eq.~(\ref{spec}), it is straightforward to translate the constraints on the amplitude in \eq{1stcons} and \eq{2ndcons} to the spectral densities,
\begin{eqnarray}
\rho^{1111}_l(s) &=& \rho^{2222}_l(s) \\
    \rho_l^{1111}(s) &=& \rho_l^{1212}(s)+\rho_l^{1221}(s) + \rho_l^{1122}(s). 
\end{eqnarray}
Now, from $tu$-crossing symmetry, we have $\rho_l^{ijkl}(s) = (-1)^l\rho^{ijlk}_l$. This implies that $\rho_l^{1111}$ and $\rho_l^{1122}$ vanish  for odd $l$ and   \eq{2ndcons} is trivially satisfied. For even $l$, on the other hand, we get the non trivial constraint, 
\begin{eqnarray}
    \rho_l^{1111} - 2\rho_l^{1212} - \rho_l^{1122} = 0.
\end{eqnarray}
As far as scattering involving $Z_L/\phi_3$ and $h$ is concerned, we can again translate the results of Sec.~\ref{amp} to constraints on the spectral densities: $\rho_l^{12ii}=\rho_l^{1ii2}=\rho_l^{1i2i}=0$, $\rho_l^{1i1i} = \rho_l^{2i2i}$,  $\rho_l^{11ii} = \rho_l^{22ii}$ and $\rho_l^{1ii1} = \rho_l^{2ii2}$ where $i=\phi_3,h$.

\paragraph{Linear programming}
 We now want to numerically compute the allowed range for $\tilde{c}^{m,n}_{ijkl}$---and thus the allowed range for HEFT WCs---by varying the  UV spectral densities \(\rho_l^{ijkl}\)  on the right hand side of \eq{iruv}. We will also ensure that   the constraints from unitarity, $st$ crossing and  $U(1)_{em}$ symmetry are respected in this process. We will treat  this as an optimization problem   that can be solved using linear programming methods. Specifically, we utilize the \texttt{scipy.optimize.linprog} \cite{scipy-optimize-linprog}  function to perform the required computations efficiently.

To facilitate this process, we discretize the UV scale \(s^{\prime}\) in \eq{iruv}, reducing the problem to a finite set of UV spectral densities \(\rho^{ijkl}_l(s^{\prime})\), which serve as the decision variables in the optimization problem. For convenience, we transform the integration variable from \(s^{\prime}\) to \(x = \Lambda^2 / s^{\prime}\) and approximate the integral over \(x\) as a finite sum where the variable \(x\) is discretized as $n/N$ with $n=1,2,\cdots,N$. Additionally, we impose a cut-off \(l_M\) on the sum over UV partial waves. With this, we are left with only a finite number of partial wave amplitude $\rho_{l}^{ijkl}(s)$s. The values of \(l_M\) and \(N\) are chosen to be sufficiently large to ensure numerical convergence in the optimization process. With these adjustments the sum-rules in \eq{iruv} take the following form,
\begin{eqnarray}
    \tilde{c}_{ijkl}^{m,n} &=& \sum_{l=0}^{\infty}(2l+1)\int_{\Lambda^2}^{\infty}\frac{ds^{\prime}}{{s^{\prime}}^{m+n+1}}\left[\rho_l^{ijkl}(s^{\prime}) + (-1)^m \rho_l^{ilkj}(s^{\prime}) \right]\sum_{p=0}^{n}L_l^p H_{m+1}^{n-p}\nonumber\\
    &\approx&\frac{1}{\Lambda^{2(m+1)}}\sum_{l=0}^{l_M}(2l+1)\sum_{n=1}^N\frac{1}{N}\left(\frac{n}{N}\right)^{m+n-1}\left(\rho_{l,n}^{ijkl} + (-1)^m \rho_{l,n}^{ilkj}\right)\sum_{p=0}^{n}L_l^p H_{m+1}^{n-p},\nonumber\\ \label{discretesumrule}
\end{eqnarray}
where  we have defined, $\rho_{l,n}^{ijkl}=\rho_{l}^{ijkl}(\Lambda^2N/n)$. 

We also discretize the null-constraints in a similar fashion. Thus we finally obtain expressions for the coefficients, $\tilde{c}^{m,n}_{ijkl}$, written as  linear combinations of discretized spectral densities $\rho_{l,n}^{ijkl}$ which are subjected to null-constraints, unitarity constraints and constraints arising from $U(1)_{em}$ symmetry.  Optimizing linear combinations of $\tilde{c}^{m,n}_{ijkl}$ over all possible variables $\rho_{l,n}^{ijkl}$ that respects a set of linear constraints, is a well-defined linear programming problem. For this work, we are only interested in the terms that are most important phenomenologically, namely, the $\tilde{c}^{2,0}_{ijkl}$. With our assumption that EFT loops can be neglected, these can be expressed as some linear combinations of WCs of HEFT operators in Table~\ref{Table1}. In Table~\ref{tab:linprog_details} we explicitly summarize the linear programming model that constrains $\tilde{c}^{2,0}_{ijkl}$s related to the $V_L V_L\to V_L V_L$ scattering amplitude.

\begin{table}[t]
    \centering
    \renewcommand{\arraystretch}{1.6} 
    \begin{tabularx}{\textwidth}{@{} >{\centering\arraybackslash}X @{}}
    \hline\hline
         \textbf{Decision Variables}  \\\hline
         
$\rho_{l,n}^{1111}, \rho_{l,n}^{3333}, \rho_{l,n}^{1212}, \rho_{l,n}^{1122}\; \rho_{l,n}^{1133}\quad\text{for }l=0,\cdots,l_M,\;\; n=1,\cdots,N$\\ \hline

         \textbf{Objective Function}  \\\hline
$\tilde{c}^{2,0}_{1111}, \tilde{c}^{2,0}_{1212}, \tilde{c}^{2,0}_{1122},\tilde{c}^{2,0}_{1111},\tilde{c}^{2,0}_{1313}\;\text{and}\;\tilde{c}^{2,0}_{1133},\quad\text{defined in \eq{discretesumrule}}$\\\hline

          \textbf{Constraints} \\\hline
          \textbf{Null Constraints} \\ 
$\sum_{l=0,\cdots,l_M}(2l+1)\sum_{n=1}^N\frac{1}{N}\left(\frac{n}{N}\right)^{r+2}C^{iiii}_{r,i_r}(l)\rho_{l,n}^{iiii}=0\quad\text{for $i=1,3$}$\\
$\sum_{l=0,\cdots,l_M}(2l+1)\sum_{n=1}^N\frac{1}{N}\left(\frac{n}{N}\right)^{r+2}\left(C^{1i1i}_{r,i_r}(l)\rho_{l,n}^{1i1i}+C^{11ii}_{r,i_r}(l)\rho_{l,n}^{11ii} \right)=0\quad\text{for $i=2,3$}$\\\hline

          \textbf{Unitarity Constraints} \\ 
$0\leq\rho_{l,n}^{iiii}\leq2 \quad\text{for $i=1,3$}$\\
$0\leq \rho_{l,n}^{1i1i}\leq1/2 \quad\text{for $i=2,3$}$\\
$-1\leq \rho_{l,n}^{11ii}\leq1 \quad\text{for $i=2,3$}$\\
$\left\vert\rho_{i,n}^{1122}\right\vert\leq1-\left\vert1-\rho_{l,n}^{1111} \right\vert$\\
$\left\vert\rho_{i,n}^{1133}\right\vert\leq1-\left\vert1-\frac{\rho_{l,n}^{1111}+\rho_{l,n}^{3333}}{2} \right\vert$\\\hline

          \textbf{Symmetry Constraints} \\ 
$\rho_{l,n}^{1111}-2\rho_{l,n}^{1212} -\rho_{l,n}^{1122} = 0$\\\hline \hline
    \end{tabularx}
    \caption{Description of decision variables, objective function, and constraints for the optimization problem in the $V_L V_L \to V_L V_L$ case. Here the $C^{ijkl}_{r,i_r}(l)\;$s are polynomials in $l$, see~\cite{Chen:2023bhu}}
    \label{tab:linprog_details}
\end{table}
 \section{Visualizing the HEFT-hedron: final positivity constraints}
 \label{hefthedron}

\begin{figure}[t!]
    \centering
    \includegraphics[width=0.85\textwidth]{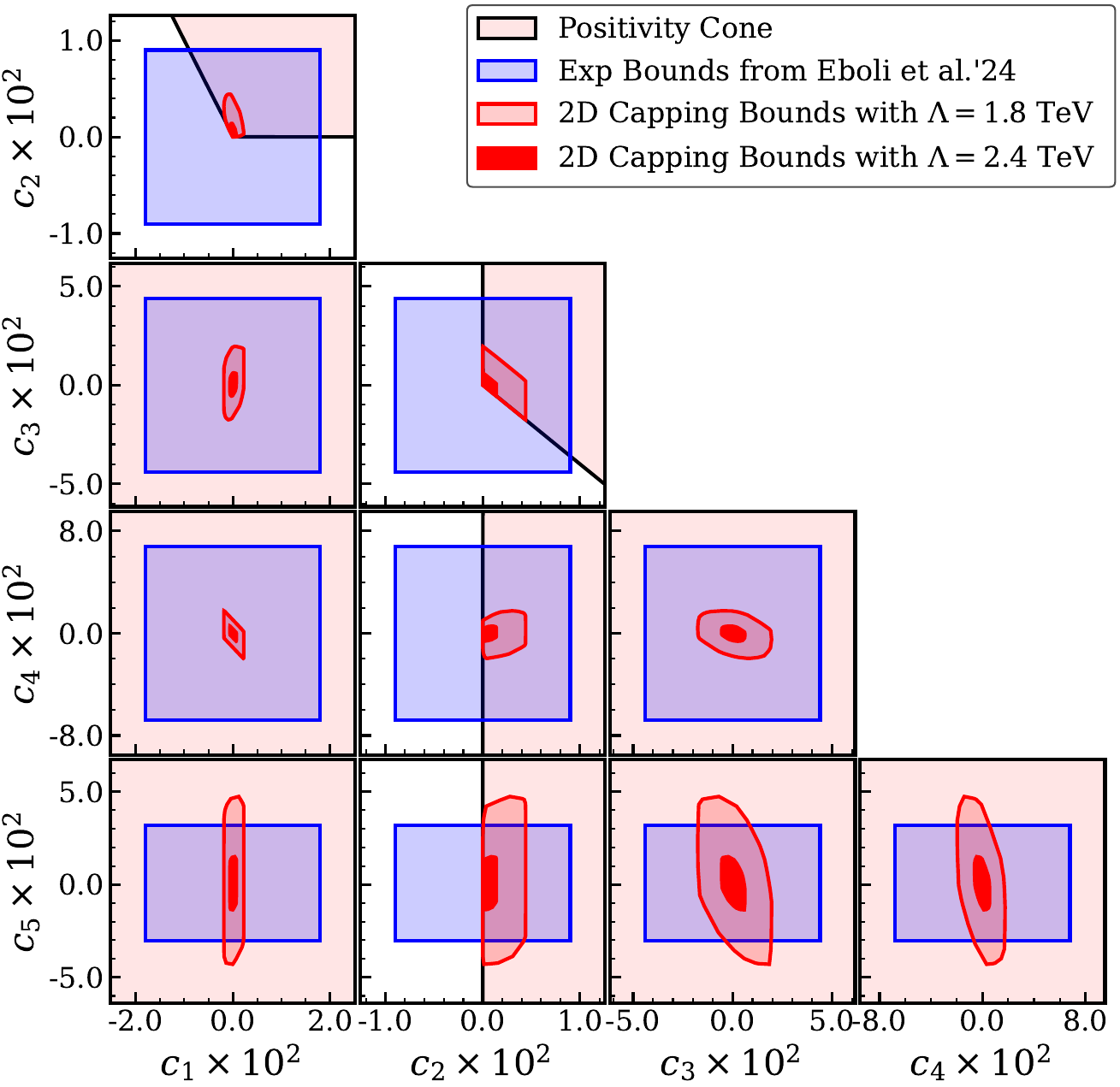}
    \caption{We show the region allowed by positivity constraints in the 5-dimensional space of the WCs, $c_1-c_5$, contributing to the  longitudinal vector boson scattering, $V_L V_L \to V_L V_L$. These constraints can be directly translated to  the space of  TGCs and aQGCs using the mapping provided in Table~\ref{heftmap}. In pink we   show the conical region allowed by the analytical constraints in Table~\ref{relations}. The solid  red  line shows the boundary of the region allowed after the double-sided bounds are imposed using the numerical  procedure detailed in Sec.~\ref{dsided} is followed taking, $\Lambda =1.8$ TeV, whereas in dark red we show the region obtained by the same procedure taking, $\Lambda =2.4$ TeV. The bounds on a particular 2-dimensional plane are obtained after marginalizing all other directions. The blue shaded region is the experimentally allowed  region  derived in Ref.~\cite{Eboli:2023mny}.  }
    \label{figvvvv}
\end{figure}


\begin{figure}[t!]
    \centering
    \includegraphics[width=0.8\textwidth]{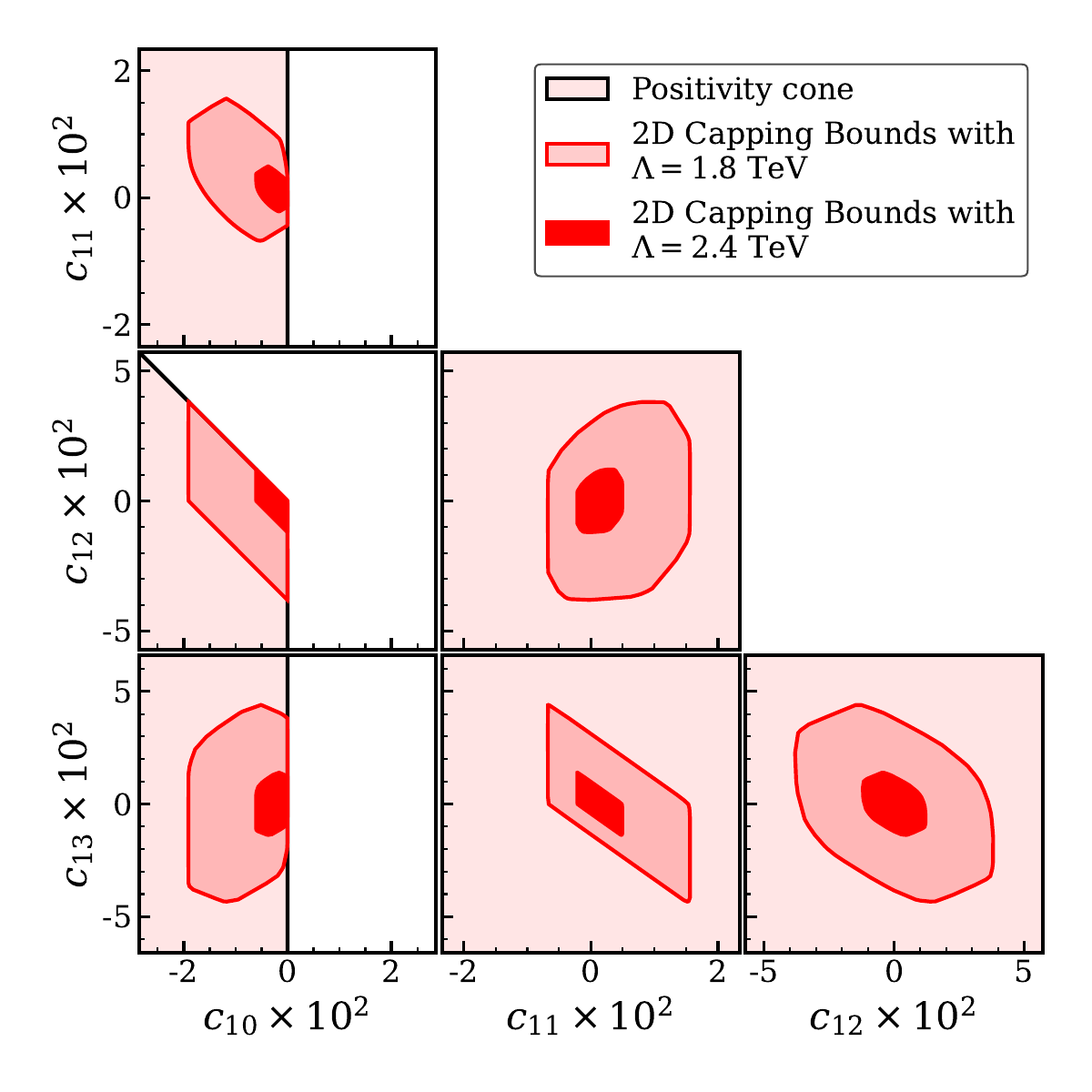}
    \caption{We show the region allowed by positivity constraints in the 5-dimensional space of the WCs, $c_{10}-c_{13}$, contributing to the    scattering, $V_L V_L \to hh$. Again, the constraints can be directly translated to  the space of anomalous  couplings using the mapping provided in Table~\ref{heftmap}. The pink region  shows the  allowed conical region derived from the analytical constraints in Table~\ref{relations}. The solid  red  line shows the boundary of the allowed region derived  using the numerical  procedure detailed in Sec.~\ref{dsided} with  $\Lambda =1.8$ TeV.  In dark red we show the region obtained by the same procedure but with  $\Lambda =2.4$ TeV. Once again, the bounds on a particular 2-dimensional plane are obtained after marginalizing all other directions. }
    \label{figvvhh}
\end{figure}

\begin{figure}[t!]
    \centering
    \includegraphics[width=0.8\textwidth]{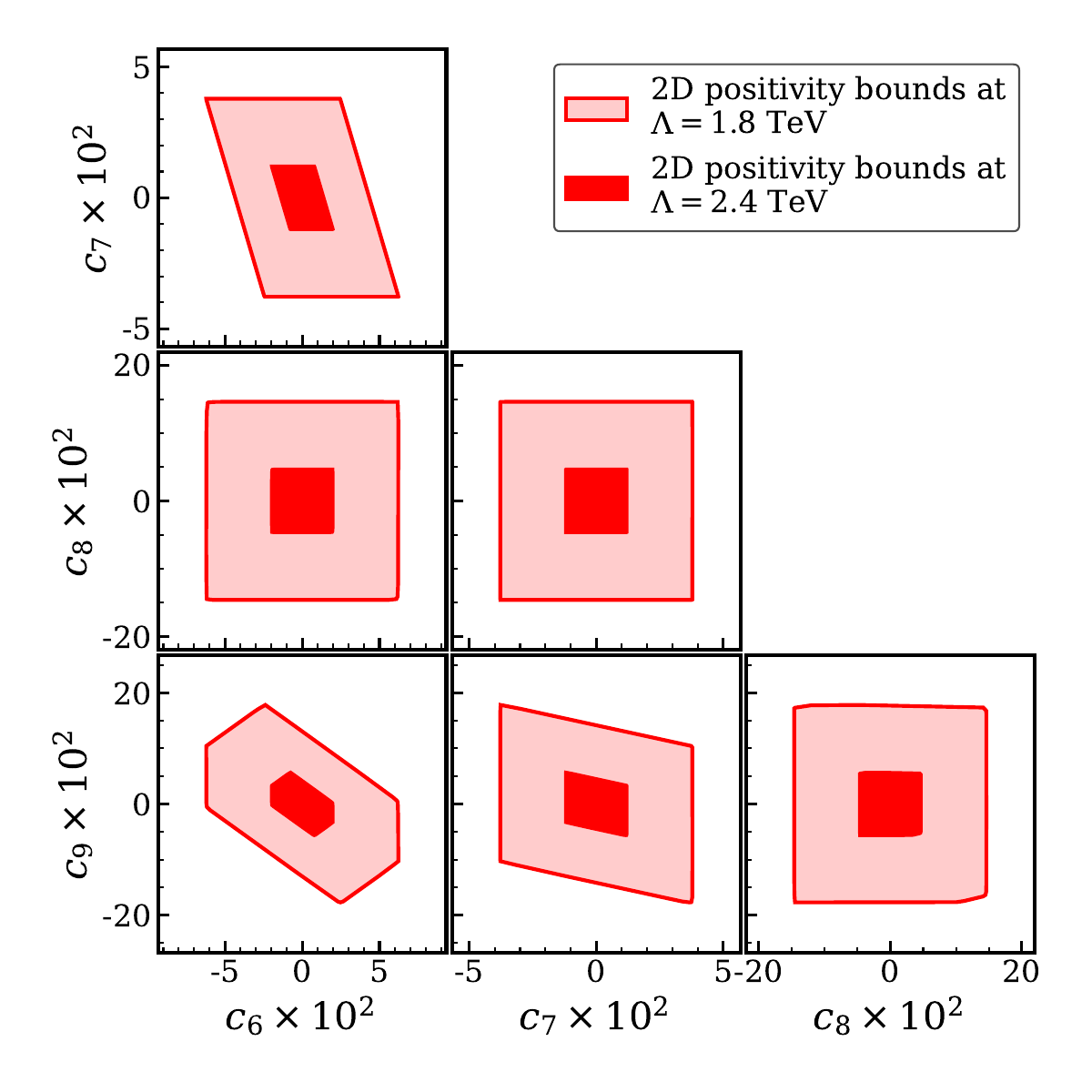}
    \caption{ Bounds on the WCs $c_6-c_8$ using the second set of positivity constraints on Table~{relations}. For the upper bound on the right hand side of these inequalities we use the numerical capping bounds in Figs.~\ref{figvvvv} and~\ref{figvvhh} and \eq{hhhh} for  $\Lambda=1.8$ TeV and 2.4 TeV. The allowed region in light and dark red, respectively, correspond to the former and latter choice of $\Lambda$.}
    \label{figvvvh}
\end{figure}


 In this section we first present our final constraints on the HEFT WCs of Table~\ref{Table1} in Sec.~\ref{r1}. We then discuss the implications of applying the SMEFT constraints from Table~\ref{tab:SMEFT_constraints} in Sec.~\ref{r2}.
 
 \subsection{Positivity constraints on HEFT WCs}
 \label{r1}
 In Fig.~\ref{figvvvv},~\ref{figvvhh} and~\ref{figvvvh} we, respectively,  show the bounds on the WCs contributing to $V_L V_L \to V_L V_L$ , $V_L V_L \to hh$ and $V_L V_L \to V_Lh$ processes. For each of the three subsets of WCs we present the projection of our bounds on all possible  two dimensional planes. In Table~\ref{tab:linear_constraints} we present the our final constraint on each individual WC   after marginalizing over all others. Table~\ref{tab:linear_constraints}  also includes bounds on $c_{14}$, the  only operator contributing to the $V_L h \to hh$ amplitude, and $c_{15}$,   the only operator contributing to the $hh \to hh$ process. Our bounds can be directly translated to the space of anomalous couplings using the mapping provided in Table~\ref{heftmap}.
 
 \subsection*{Constraints on WCs contributing to $V_LV_L \to V_L V_L$}
 First, we present the bounds on the WCs contributing to longitudinal vector boson scattering, $V_LV_L \to V_L V_L$ process, i.e the set of 5  WCs, $\{c_1, c_2 \cdots c_5\}$, in Fig.~\ref{figvvvv}.  We show the positivity constraints obtained in  this work in different shades of red whereas the experimental bounds have been shown in blue. The pink region shows the region allowed after applying the analytical constraints in Table~\ref{relations}---in particular the first 4 constraints from Set 1 and the first inequality from Set 2.  After application of these constraints from Table~\ref{relations}, we find a convex conical allowed region that occupies less than 26 $\%$ of  5-dimensional region allowed by the experimental bounds shown in blue.   The red boundary shows the allowed region obtained  after carrying out the numerical capping  procedure described in Sec.~\ref{dsided}; the lighter (deeper) shade of red within this boundary corresponds to the choice  $\Lambda=1.8$ TeV ($\Lambda=2.4$ TeV). The shaded regions on a particular 2-dimensional plane show the projections of the 15-dimensional HEFT-hedron on it--i.e., we marginalize over all other 13 directions while providing the bounds. It can be seen that our final bounds  are  significantly stronger than  the experimental bounds.

As far as the blue region in Fig.~\ref{figvvvv} is concerned, it shows the allowed region  after imposing the  1-dimensional experimental bounds (see Table~\ref{tab:linear_constraints}) obtained after marginalizing over all other WCs. These experimental bounds  have been derived in \cite{Eboli:2023mny} using the results from  vector boson scattering searches at CMS and ATLAS \cite{2021135992,2020135710,Aad2024}. Using our power-counting scheme in \eq{pc}, the numerical values of the experimental bounds from Table~\ref{tab:linear_constraints} can be converted to lower bounds on the new physics scale that range from $\Lambda>0.9$ TeV to $2.6$ TeV (see Table~\ref{tab:linear_constraints}).  The values of   $\Lambda$ (1.8, 2.4 TeV)   used to obtain the numerical capping  bounds inside the red solid line have  been chosen to be in the same ballpark as these experimental lower bounds.


.  
\subsection*{Constraints on WCs contributing to $V_LV_L\to hh$}
We now present the bounds on the WCs contributing to  the, $V_LV_L \to hh$ process, i.e the set of 4  WCs, $\{c_{10}, c_{11} \cdots c_{13}\}$, in Fig.~\ref{figvvhh}. Once again in pink we show the region allowed after applying the analytical constraints in Table~\ref{relations} (specifically the 5th and 6th  relations from the first set). The red boundary has been obtained  numerically using the methods of Sec.~\ref{dsided} taking  $\Lambda=1.8$ TeV; the deeper shade of red within region corresponds to taking $\Lambda=2.4$ TeV.  Once again the bounds on a particular 2-dimensional plane are obtained after marginalising all other directions.

\begin{table}[t]
\centering
\renewcommand{\arraystretch}{1.4} 
\begin{tabular}{@{} >{}c<{}  >{}c<{} >{}c<{}  >{}c<{} >{}c<{}  >{}c<{} @{}} 
\hline\hline
\multirow{2}{*}{\text{Process}} & \multirow{2}{*}{\text{Coefficients}} &\multicolumn{2}{c}{\text{Exp Bounds}} & \multicolumn{2}{c}{\text{Capping bounds$\times 10^2$}} \\ 
\cline{3-6}
& & $c_i\times 10^2$ & $\Lambda~$(TeV) & $\Lambda = 1.8$ TeV & $\Lambda = 2.4$ TeV \\\hline

\multirow{5}{*}{$V_LV_L \to V_LV_L$} & $c_1$ &  $\left[-1.8, 1.8\right]$ & $1.8$ &   $\left[-0.18,0.22\right]$         &    $\left[-0.06,0.07\right]$        \\
& $c_2$ & $\left[-0.9,0.9\right]$ & $2.6$ & $\left[0,0.44\right]$    &  $\left[0,0.14\right]$   \\ 
& $c_3$ & $\left[-4.4, 4.4\right]$ & $1.2$ & $\left[-1.76,1.95\right]$  &  $\left[-0.56,0.62\right]$   \\
& $c_4$ & $\left[-6.8,6.8\right]$  &  $0.9$ & $\left[-1.94,1.75\right]$   &  $\left[-0.61,0.55\right]$     \\
& $c_5$ & $\left[-3.0, 3.2\right]$ & $1.4$ &  $\left[-4.31, 4.78\right]]$  &  $\left[-1.36,1.51\right]$    \\ \hline
\multirow{4}{*}{$V_LV_L\to V_Lh$} & $c_{6}$ & $-$ & $-$ & $\left[-6.20,6.20\right]$   &  $\left[-1.96,1.96\right]$    \\
& $c_7$ & $-$ & $-$  & $\left[-3.78,3.78\right]$    &  $\left[-1.20,1.20\right]$   \\
& $c_8$ & $-$ & $-$  & $\left[-14.60, 14.60\right]$   &  $\left[-4.62,4.62\right]$   \\
& $c_9$ & $-$  & $-$ & $\left[-17.78,17.78\right]$  &  $\left[-5.62,5.62\right]$   \\ \hline
\multirow{4}{*}{$V_LV_L\to hh$} & $c_{10}$ & $-$ & $-$  &  $\left[-1.91,0\right]$  &  $\left[-0.60,0\right]$    \\
& $c_{11}$ & $-$ & $-$  &  $\left[-0.68,1.56\right]$   &  $\left[-0.21,0.49\right]$  \\
& $c_{12}$ & $-$ & $-$  &  $\left[-3.81, 3.81\right]$   &  $\left[-1.20,1.20\right]$  \\
& $c_{13}$ & $-$  & $-$ &  $\left[-4.40,4.40\right]$   &   $\left[-1.39,1.39\right]$   \\ \hline
$V_Lh\to hh$ & $c_{14}$ & $-$ & $-$ & $\left[-11.24,11.24\right]$   &  $\left[-3.55,3.55\right]$    \\\hline
$hh\to hh$ & $c_{15}$ & $-$ & $-$ & $\left[0,2.07\right]$   &  $\left[0, 0.66\right]$   \\\hline\hline
\end{tabular}
\caption{Experimental and positivity  bounds on the WCs of HEFT operators  in Table~\ref{Table1}. The third column presents experimental bounds on the WCs as reported in \cite{Eboli:2023mny}. The fourth and fifth columns provide the positivity  bounds which incorporate both the analytical constraints of Table~\ref{relations}   and the numerical capping bounds for two choices of the new physics scale ($\Lambda$): $1.8$ TeV and $2.4$ TeV. For both the experimental and positivity  bounds, the final constraint on each individual WC has been obtained after marginalizing over all others.}
\label{tab:linear_constraints}
\end{table}

\subsection*{Constraints on WCs contributing to $V_LV_L\to  V_L h$}

In Fig.~\ref{figvvvh} we show bounds on the WCs contributing to the $V_LV_L \to  V_L h$ process, namely on the set, $\{c_6, c_7 \cdots c_9\}$. Unlike the previous figures, the red boundary  has not been obtained by directly applying the numerical procedure of Sec.~\ref{dsided} on $c_6-c_9$. Instead we have the second set of inequalities in Table~\ref{relations} (specifically the second, third, fifth, sixth and eighth  relations of  Set 2) to put an upper bound on $c_6-c_9$ using upper bounds on the other WCs which appear on the right hand side of these inequalities and contribute to the processes, $V_LV_L \to  V_L V_L, hh$  and $hh \to hh$ processes. For the upper bounds on the latter set of  WCs, we do use the numerical bounds   for $\Lambda=1.8$ TeV and $2.4$ TeV (shown in Fig.~\ref{figvvvv} and Fig.~\ref{figvvhh}) that respectively result in the allowed region shown in light (dark) red in Fig.~\ref{figvvvh}. In the future if  experimental bounds  on the WCs contributing to the $V_LV_L \to  V_L V_L, hh$ processes become stronger than the capping ones,  we can use them instead to bound the right hand side of the inequalities in Table~\ref{relations}.

\begin{figure}[htpb!]
    \centering
    \includegraphics[width=0.75\linewidth]{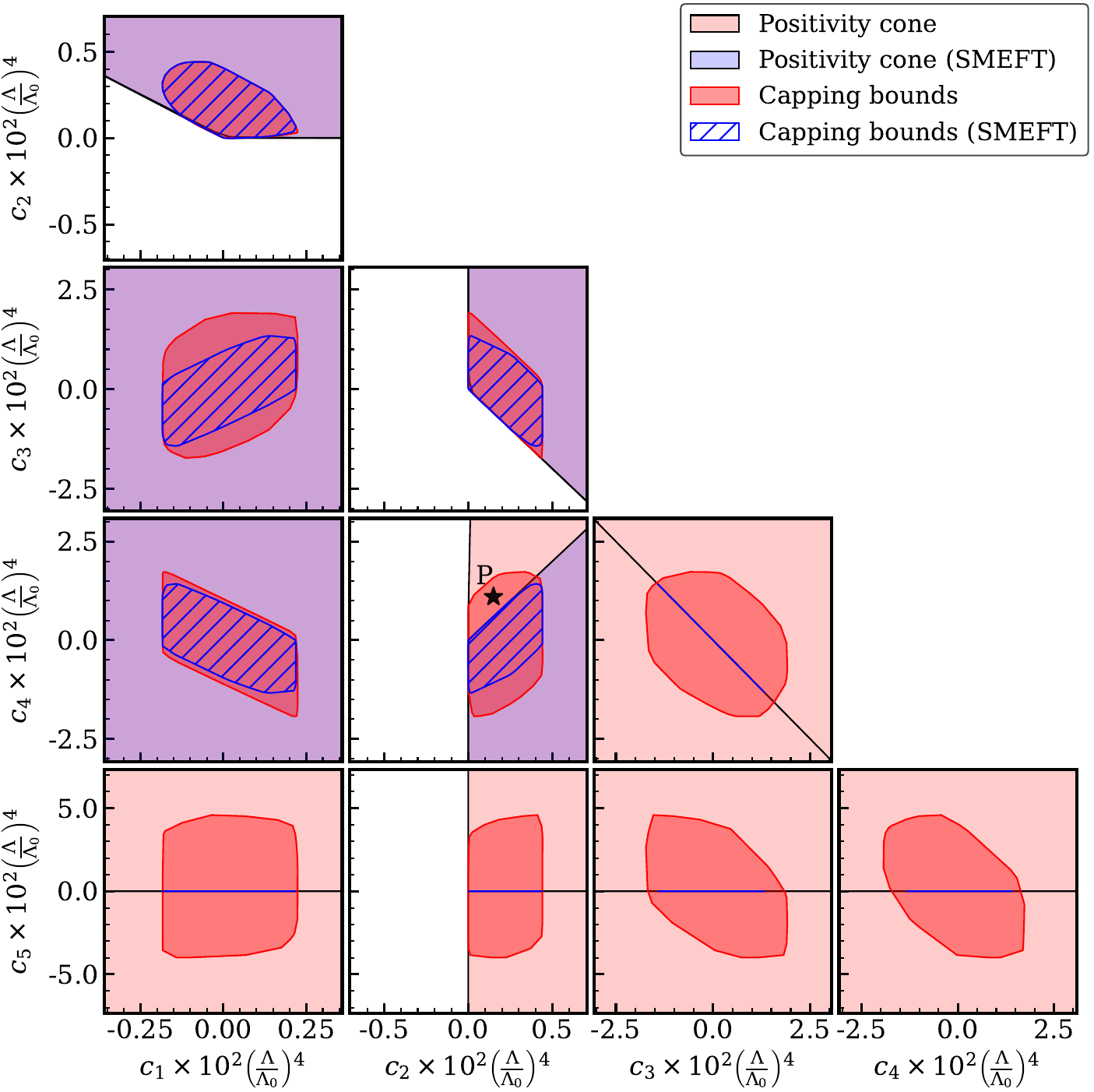}
    \caption{This figure compares the regions allowed by positivity constraints for NLO HEFT operators  contributing to longitudinal vector boson scattering, \( V_L V_L \to V_L V_L \), before and after imposing the requirements of the dimension-8 SMEFT (see Table~\ref{tab:SMEFT_constraints}). The pink region represents the   bounds for HEFT, derived using the constraints listed in Table~\ref{relations}, while the darker red region shows the capping bounds for HEFT, obtained through the numerical procedure described in Sec.~\ref{dsided}. The light blue  region (the blue hatched region) represents the regions  consistent with the analytical positivity (numerical capping) bounds as well as the SMEFT requirements of Table~\ref{tab:SMEFT_constraints}. The point, $P$, represents a measurement that lies outside the  light blue allowed region in SMEFT but within the pink region consistent with the HEFT positivity cone. Here, we have taken, $\Lambda_0 = 1.8$ TeV.}
    \label{fig:VVVV_smeft_vs_heft}
\end{figure}

\begin{figure}[htpb!]
    \centering
    \includegraphics[width=0.7\linewidth]{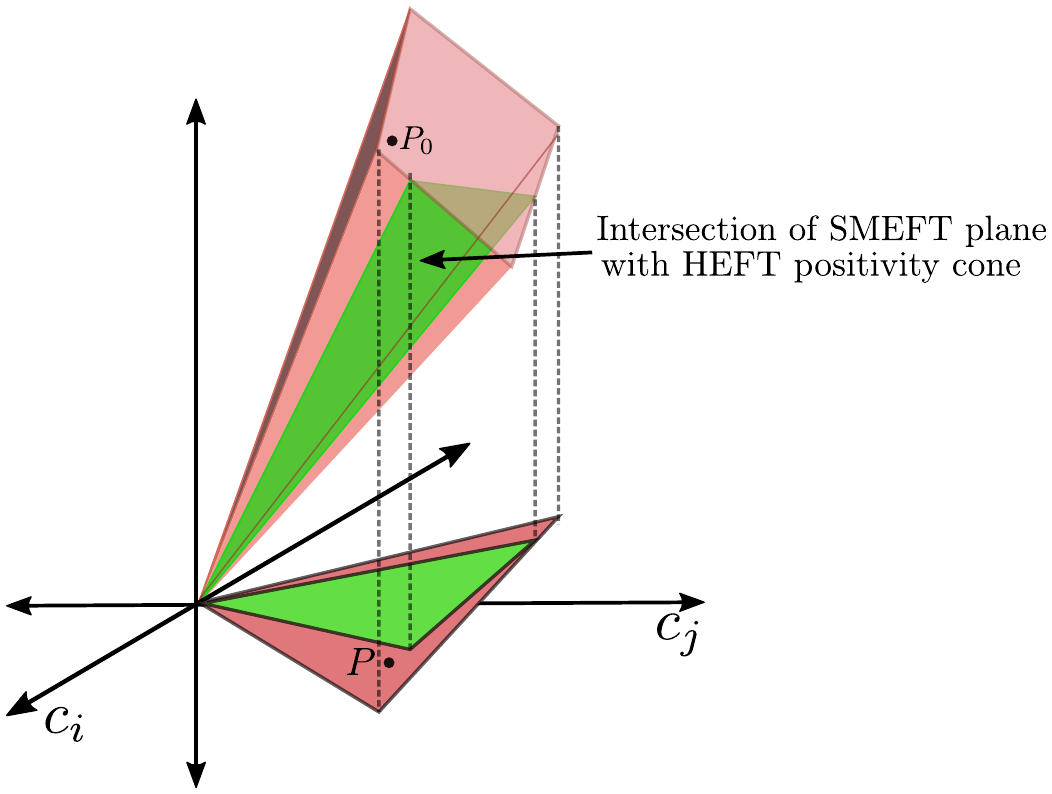}
    \caption{A schematic diagram showing HEFT positivity cone (pink) and its intersection with SMEFT plane (green) and the projection of these two regions on the space of the 2 HEFT WCs $c_i$ and $c_j$.}
    \label{fig:schematic}
\end{figure}
\subsection*{Constraints on WCs contributing to $hh \to  V_L h,hh$}

Finally, let us provide constraints on $c_{14}$, which is the only WC contributing to the $VV \to Vh$ process, and $c_{15}$, the only WC contributing  to the $hh \to  hh$ process. We obtain, 
\begin{eqnarray}
    &\left\vert c_{14}\right\vert\left(\frac{\Lambda}{1.8 \text{ TeV}}\right)^4& < 0.112  \\
    0 < &c_{15}\left(\frac{\Lambda}{1.8 \text{ TeV}}\right)^4& < 0.021
    \label{hhhh}
\end{eqnarray}
where the positivity of $c_{15}$ follows from the analytical method described in  Sec.~\ref{positivity} and the upper bounds on the magnitude of $c_{14}$ and $c_{15}$  have been obtained following the procedure of Sec.~\ref{dsided}.

\subsection{Positivity constraints in the SMEFT}
\label{r2}

In this section we  will obtain the regions in the HEFT parameter space that are consistent with both positivity constraints as well as the SMEFT at dimension-8 level, i.e. they obey the  constraints of Table~\ref{tab:linear_constraints}. We will consider the case of WCs contributing to $V_L V_L \to V_L V_L$ in this section and discuss similar analyses for the $V_L V_L \to hh, V_L h$ processes  later in Appendix~\ref{smeft}. In Fig.~\ref{fig:VVVV_smeft_vs_heft} we show, in blue, the regions that are consistent with the SMEFT constraints in Table~\ref{tab:SMEFT_constraints} as well as the analytical positivity constraints of Table~\ref{relations}. The blue hatched region in Fig.~\ref{fig:VVVV_smeft_vs_heft}  represents the region consistent with both the  SMEFT constraints and the numerical capping bounds.\footnote{Note that as far as the numerical capping bounds are concerned, we    expect them to be less stringent than the bounds obtained in  Ref.~\cite{Chen:2023bhu, Hong:2024fbl}.  This is because  Ref.~\cite{Chen:2023bhu, Hong:2024fbl} used the full $SU(2)_L \times U(1)_Y$ symmetry to constrain the spectral densities to obtain the double-sided bounds (see Sec.~\ref{dsided}). Our assumption, that the spectral densities satisfy $U(1)_{\text{em}}$ invariance, is thus less restrictive than that of Ref.~\cite{Chen:2023bhu, Hong:2024fbl}.}   Geometrically these regions represent the projection of---the intersection  of the the 15  dimensional region allowed by positivity constraints with the 3 dimensional SMEFT hyperplane---on the different 2-dimensional HEFT planes  in Fig.~\ref{fig:VVVV_smeft_vs_heft}. For the case of the analytical positivity constraints,  we show this schematically  in  Fig.~\ref{fig:schematic}. Recall that the SMEFT positivity cone, i.e the intersection of the SMEFT plane and the HEFT positivity cone, was derived in \eq{smeftcone}.

The regions in Fig.~\ref{fig:VVVV_smeft_vs_heft} in pink  (red)  represent  the region allowed in the HEFT by analytical (numerical) positivity bounds. They represent the projection of the full 15-dimensional allowed region---and not just the intersection with the SMEFT plane---on the various 2-dimensional planes of Fig.~\ref{fig:VVVV_smeft_vs_heft}. As shown in Fog.~\ref{fig:schematic}, we expect them to be generally larger than the corresponding blue (hatched blue) region  consistent with SMEFT.   This fact has important implications that were recently pointed out in  Ref.~\cite{Rodd}. Consider, for instance, the scenario where HEFT is the correct EFT to describe nature  and a non-zero measurement, $P$, is made in the $c_2-c_4$ plane such that is outside the region consistent with the SMEFT positivity cone,  but within the region consistent with the HEFT positivity cone (see Fig.~\ref{fig:VVVV_smeft_vs_heft}). If one wrongly assumes that SMEFT is the correct EFT at low-energies, such a  measurement would  not necessarily be indicative of the breakdown of unitarity or causality but might only mean that the correct low energy EFT is  not the  SMEFT, but a more general HEFT. \footnote{Note that if $P$ was only outside the  region allowed by capping but not outside the region consistent with the positivity cone, it would be more difficult to draw any of these inferences. This is because with the choice of a smaller, $\Lambda$, the region allowed by the capping bounds would become larger and might start including the  point, $P$.  Strong conclusions can be drawn in this case only if experiments definitively exclude any possibility of new physics below the chosen cutoff scale, $\Lambda$. }


As shown in Fig.~\ref{fig:schematic}, this possibility can arise only if the correct theory of nature corresponds to a point, $P_0$,  in the 15-dimensional HEFT space that lies outside the  SMEFT plane. Its projection, $P$,  on the $c_2-c_4$ plane would then lie outside the blue region. This would still not be very surprising if all the HEFT WCs, $c_1-c_{15}$ can be independently measured---recall that  there is a one-to-one mapping between these HEFT WCs and the  the forward amplitude for different channels of longitudinal gauge-Higgs scattering (see \eq{c20}). This is because one could then directly check that the SMEFT constraints in Table~\ref{tab:SMEFT_constraints}
are violated---that is   the theory lies  outside the SMEFT plane. For such a scenario, the positivity conditions of Table~\ref{relations} would add no new information over what is already contained in Table~\ref{tab:SMEFT_constraints}.

In practice, however, not all of the HEFT WCs, $c_1-c_{15}$, would  be measurable. The observation made in Ref.~\cite{Rodd} can then become truly powerful. Consider, for instance, the hypothetical scenario where future experiments are sensitive only to the HEFT WCs $c_2$ and $c_4$. As all values of $c_2$ and $c_4$ are consistent with requirements of Table~\ref{tab:SMEFT_constraints}, there would be no way to use these requirements to infer that SMEFT is the wrong  low-energy description. Measurements converging at the point $P$ outside the SMEFT positivity region might then provide the first indication that SMEFT is not the correct EFT description.

Finally, let us discuss another subtlety that may arise if one wishes to utilize the proposal of Ref.~\cite{Rodd}. Ref.~\cite{Rodd} demonstrated that taking only the 5 custodial invariant operators of Table~\ref{Table1}, the projection of the HEFT positivity cone on  the SMEFT plane  is larger than than their intersection region.  When one projects to a certain  set of observables, however, this feature may be lost. For instance, let us consider the possibility of utilizing  the proposal of Ref.~\cite{Rodd} in vector boson scattering measurements. We must then project to the plane of the only custodial invariant operators contributing to this process, namely the   $c_1-c_2$ plane. We see in Fig.~\ref{fig:VVVV_smeft_vs_heft} that the regions allowed by SMEFT and HEFT positivity constraints in this plane are, unfortunately, identical. One can, in fact, explicitly check that if the value of  SMEFT WCs is fixed by a measurement of only $c_1$ and $c_2$, and if these two WCs satisfy the HEFT positivity constraints in Table~\ref{relations}, the inferred SMEFT WCs will always be in the SMEFT positivity cone defined by  \eq{smeftcone}. Thus in the custodial limit, there is no way to  distinguish SMEFT from HEFT using aQGC measurements alone. If we consider the $V_L V_L \to hh$ process, however, the relevant plane of custodial invariant operators  is  the $c_{10}-c_{11}$ plane and the region allowed in HEFT is then clearly larger than that in SMEFT, see Fig.~\ref{fig:VVhh_smeft_vs_heft}.

We discuss in  the implications of the SMEFT constraints of Table~\ref{tab:SMEFT_constraints} on the $V_L V_L \to V_L V_L, V_L h$ process in Appendix~\ref{smeft}.

\section{Conclusions}
\label{conclusions}

We derive the consequences of causality/analyticity and unitarity on the classic problem of scattering among  longitudinal electroweak and Higgs bosons. In particular, we derive the constraints from these theoretical requirements on the the coefficient of $s^{2}$ in the forward amplitude for these processes. Using only  $U(1)_{em}$ invariance, we show that the $s^2$ piece of the forward amplitudes of  longitudinal gauge-Higgs scattering processes can be parameterized by 15 independent parameters (see Sec.~\ref{amp}). We also provide a one-to one mapping of these 15 parameters with 15 WCs of the NLO HEFT lagrangian (see \eq{c20}) and 15 independent linear combinations of anomalous couplings of a $U(1)_{em}$ lagrangian (see Table~\ref{heftmap}). As far as the HEFT is concerned, no other scattering process grows as $s^2$ or faster at NLO, so that considering only these 15 operators allows us to obtain the complete set of bounds at NLO. In the SMEFT only three operators contribute to the $s^2$ piece of the forward amplitudes up to dimension-8 level. Thus the SMEFT truncated at dimension-8 level implies 12 constraints on the space of HEFT WCs, that  we present in Table~\ref{tab:SMEFT_constraints}.

We then   derive analytical positivity constraints on the space of these 15 HEFT WCs. These have been presented in Table~\ref{relations} in two sets. The first set of constraints requires the positivity of some linear combinations of CP-even WCs contributing to the elastic processes, $V_Lh \to V_Lh,~Z_LZ_L \to Z_LZ_L, W_LW_L \to W_LW_L, W_LZ_L \to W_LZ_L$ and $hh\to hh$. The second set of constraints,  imply that the magnitude of certain WCs---including all CP-odd ones---that contribute to the inelastic processes, $V_LV_L \to h V_L, hh \to V_L h$, must be smaller than products of the WCs constrained by the first set. A non-vanishing WC contributing to the $V_LV_L \to h V_L, hh \to V_L h$ processes, would thus imply that some of the other WCs contributing to the elastic processes must be non-zero. Together these analytical constraints define a positivity cone within which the HEFT WCs must lie. They rule out about 95 $\%$ of the full 15-dimensional HEFT space and about  about 74 $\%$ of the 5 dimensional space still allowed by  experimental bounds from vector boson scattering processes.

We then go on to derive numerical double-sided bounds on the HEFT WCs,  that cap this positivity cone. Our final results,  shown in Fig.~\ref{figvvvv}-\ref{figvvvh} and Table~\ref{tab:linear_constraints}, provide the first reported bound on the WCs contributing to the $V_L V_L  \to V_L h, hh$ and $hh  \to V_L h, hh$  processes. For the case of vector boson scattering, they significantly improve over existing experimental bounds.  Finally, we obtain the region  in HEFT space allowed by positivity  as well as the requirements of the dimension-8 SMEFT.  We then comment on the possibility~\cite{Rodd} of using positivity bounds to infer that the HEFT---and not the SMEFT---is the correct low energy EFT.

The LHC has only begun to probe the scattering of gauge and Higgs bosons---a set of processes of great conceptual importance. With the large increase in integrated luminosity expected  in the coming decades, it will be able to probe the 15-dimensional space of EFT WCs discussed here with greater and greater  precision. Our results   provide  theoretical priors on this space to complement this important experimental program.



\section*{Acknowledgements}
We are grateful to Amol Dighe for insightful discussions and suggestions on how to visually present our multidimensional bounds. We also thank Avik Banerjee, Shankha Banerjee and Siddhartha Karmakar for  useful discussions. We acknowledge the support from the Department of Atomic Energy (DAE), Government of India, under Project Identification Number RTI 4002.

\appendix
 \section{Vector boson scattering in the unitary gauge}
 \label{appA}
 In this section, we explicitly show that the operators in \eq{oblique} (\eq{obliquesmeft}) in the HEFT (SMEFT) do not give rise to $s^2$ growth in the vector boson scattering amplitude. 
The $s^2$ piece in the  scattering amplitudes of longitudinal gauge bosons $W^{\pm}_L,Z_L$, contains not only contributions from the aQGCs but also from the TGCs as noted in the main text (see Section \ref{anom}).  When we take into account all the TGCs and aQGCs generated by \eq{oblique} and \eq{obliquesmeft}, however, their contributions to the $s^2$ term cancels out.

First of all, for the process $Z_L Z_L \to Z_L Z_L$,   the only anomalous coupling that contributes is the aQGC,  $h^Q_{ZZ}$; this aQGC receives no contribution from the operators in \eq{oblique} (\eq{obliquesmeft})  in the HEFT (SMEFT). For the other processes the amplitude is given by, 
\begin{eqnarray}\label{vbshi}
\mathcal{M}_{W^+W^- \to W^+W^-}(s,t)&=& -\left(\frac{g^2 \ctw^2}{2 M_W^4} \delta \kappa^Z  +\frac{g^2 \stw^2}{2 M_W^4} \delta \kappa_\gamma \right)(s^2+4st+t^2) \nonumber\\ &&\quad\quad+\frac{g^2}{4 M_W^4}\left(2\delta g^Q_{WW1}(s+t)^2 - \delta g^Q_{WW2}(s^2+t^2)\right) \nonumber\\
\mathcal{M}_{W^+Z \to W^+Z}(s,t)&=& -\frac{g^2 \ctw^2}{4m_W^2 m_Z^2}\Bigg(2\delta g^Z_1(2s^2+2st-t^2) -\delta g^Q_{ZZ1} (2s^2+2st+t^2) \nonumber\\ &+& 2\delta g^Q_{ZZ2}t^2 \Bigg).
\end{eqnarray}
Here we have retained only the terms which grow quadratically with energy. All other amplitudes involving four longitudinal gauge bosons can be derived from the above two amplitudes via crossing symmetry relations.
We can explicitly check from eq.~(\ref{tgc}-\ref{qgc}) in the HEFT, and eq.~(\ref{tgcsmeft}-\ref{gcsmeft}) in the SMEFT that the   amplitude in \eq{vbshi} has no dependence on the WCs of the operators in question.

\section{Derivation of Positivity Constraints on HEFT Operators at NLO}\label{beta}
In this section, we provide a detailed discussion of how we derived the positivity constraints presented in Table~\ref{relations}. As outlined in eq.~(\ref{pos_cond}), we exploit the fact that the matrix \(\gamma_{\bm{\beta}}\), defined as
\begin{eqnarray}
\left(\gamma_{\bm{\beta}}\right)_{ik}= \beta_{j}\beta^{\ast}_{l}\frac{\partial^2}{\partial s^2}\tilde{\mathcal{M}}_{ijkl}(s,t\to 0)\vert_{s=0},
\end{eqnarray}
is positive-definite for any arbitrary choice of \(\bm{\beta}\), provided that the norm \(\sum_{i=1}^4 \left|\beta_i\right|^2 = 1\). In this expression, the indices \(i\), \(j\), \(k\), and \(l\) in \(\tilde{\mathcal{M}}_{ijkl}\) take values from 1 to 4, where \(1\), \(2\), and \(3\) correspond to the three Goldstone bosons arising from electroweak symmetry breaking, while \(4\) denotes the physical Higgs field \(h\). We can express each element of the $4\times4$ matrix $\gamma_{\bm{\beta}}$  as some linear combination of WCs of HEFT NLO operators listed in Table~\ref{Table1} and $\bm{\beta}$ which we provide below,
\begin{eqnarray}
    \left(\gamma_{\bm{\beta}}\right)_{11} &=& a_1\left\vert \beta_1\right\vert^2 + a_4\left\vert\beta_2\right\vert^2 + a_6\left\vert\beta_3\right\vert^2 + a_{11}\left\vert\beta_4\right\vert^2 + 2a_9\Re{\beta_3\overline{\beta_4}} \label{4cross4_11}\\
    \left(\gamma_{\bm{\beta}}\right)_{22} &=& a_{4}\left\vert \beta_1\right\vert^2 + a_{1}\left\vert\beta_2\right\vert^2 + a_{6}\left\vert\beta_3\right\vert^2 + a_{11}\left\vert\beta_4\right\vert^2 + 2a_{9}\Re{\beta_3\overline{\beta_4}} \\
    \left(\gamma_{\bm{\beta}}\right)_{33} &=& a_{6}\left(\left\vert \beta_1\right\vert^2 + \left\vert\beta_2\right\vert^2\right) + a_5\left\vert\beta_3\right\vert^2 + a_{13}\left\vert\beta_4\right\vert^2 + 2a_{12}\Re{\beta_3\overline{\beta_4}} \\
    \left(\gamma_{\bm{\beta}}\right)_{44} &=& a_{11}\left(\left\vert \beta_1\right\vert^2 + \left\vert\beta_2\right\vert^2\right) + a_{13}\left\vert\beta_3\right\vert^2 + a_{16}\left\vert\beta_4\right\vert^2 + 2a_{15}\Re{\beta_3\overline{\beta_4}} \\
    \left(\gamma_{\bm{\beta}}\right)_{12} &=& 2a_{2}\Re{\beta_1\overline{\beta_2}} \\
    \left(\gamma_{\bm{\beta}}\right)_{13} &=& 2a_{3}\Re{\beta_1\overline{\beta_3}} + 2a_{7}\Re{\beta_1\overline{\beta_4}} + 2a_{10}\Re{\beta_2\overline{\beta_4}} \\
    \left(\gamma_{\bm{\beta}}\right)_{14} &=& 2a_{7}\Re{\beta_1\overline{\beta_3}} + 
    2a_{8}\Re{\beta_1\overline{\beta_4}} - 2a_{10}\Re{\beta_2\overline{\beta_3}} \\
    \left(\gamma_{\bm{\beta}}\right)_{23} &=& -2a_{10}\Re{\beta_1\overline{\beta_4}} + 2a_{3}\Re{\beta_2\overline{\beta_3}}+ 2a_{7}\Re{\beta_2\overline{\beta_4}} \\
    \left(\gamma_{\bm{\beta}}\right)_{24} &=& 2a_{10}\Re{\beta_1\overline{\beta_3}} + 2a_{7}\Re{\beta_2\overline{\beta_3}}+ 2a_{8}\Re{\beta_2\overline{\beta_4}} \\
    \left(\gamma_{\bm{\beta}}\right)_{34} &=& a_{9}\left(\left\vert \beta_1 \right\vert^2 + \left\vert \beta_2 \right\vert^2\right) +  a_{12}\vert\beta_3\vert^2 + a_{15}\vert\beta_4\vert^2 + 2a_{14}\Re{\beta_3\overline{\beta_4}}\label{4cross4_34}
\end{eqnarray}
Given that, the matrix $\gamma_{\bm{\beta}}$ is hermitian,  we can express the remaining elements in terms of those already provided above. Coefficients $a_i$'s in Eq.(\ref{4cross4_11}-\ref{4cross4_34}) are related to the WCs $c_i$'s in Eq.~(\ref{HEFT_WCs}) in the following manner,
\begin{eqnarray}
a_1&=&16(c_1+c_2),\quad a_2=4(2c_1+c_2),\quad a_3=8c_1 + 2c_2 + c_3 + 4c_4,\nonumber\\
a_4&=&8c_2, \quad a_5=16(c_1+c_2)+8(c_3+c_4)+4c_5,\quad a_6=2(4c_2+c_3),\nonumber\\
a_7&=&c_6+\frac{c_7}{2},\quad a_8=2c_{11},\quad a_9=c_7, \quad a_{10}=-\frac{c_8}{4},\quad a_{11}=-2c_{10},\\
a_{12}&=&2(c_6+c_7+c_9/2),\quad a_{13}=-(2c_{10}+c_{12}),\quad a_{14}=(2c_{11}+c_{13}),\nonumber\\
&\quad&\hspace{2cm}a_{15}=-c_{14}/2,\quad a_{16}=4c_{15}.\nonumber
\end{eqnarray}

We intend to derive conditions on $a_i~$s so that matrix $\gamma_{\bm{\beta}}$ satisfies positive definiteness condition for an arbitrary choice of $\bm{\beta}$. To this end, we utilize the property that any principle sub-matrix of a positive definite matrix is also positive definite. First, we demand all $1\times1$ principle sub-matrices, i.e.   the diagonal elements of $\gamma_{\bm{\beta}}$, are positive. For example, for different  choices for $\bm{\beta}$, $\left(\gamma_{\bm{\beta}}\right)_{11}>0$ implies,
\begin{eqnarray}
    a_1&>&0 \hspace{1.2cm}\textit{$\beta_1\neq0$ and $\beta_i=0$ for $i=2,3,4$}\\
    a_4&>&0 \hspace{1.2cm}\textit{$\beta_2\neq0$ and $\beta_i=0$ for $i=1,3,4$}\\
    a_6&>&0 \hspace{1.2cm}\textit{$\beta_3\neq0$ and $\beta_i=0$ for $i=1,2,4$}\\
    a_{11}&>&0 \hspace{1.2cm}\textit{$\beta_4\neq0$ {\rm and} $\beta_i=0$ for $i=1,2,3$}\\
    a_9^2&<&a_6a_{11}\hspace{0.2cm}\textit{$\arg \beta_3 = \pi + \arg \beta_4 $,  $\vert \beta_3\vert/\vert \beta_4\vert = \sqrt{a_{11}/a_6}$ and $\beta_{1,2}=0$. }
\end{eqnarray}
Following similar steps for rest of the diagonal elements of $\gamma_{\bm{\beta}}$, we get the following set of constraints,
\begin{eqnarray}
    &a_1,a_4,a_6,a_{11},a_{5},a_{13},a_{16}>0,\\
    &a_9^2<a_6a_{11},\;\;a_{12}^2<a_5a_{13},\;\;a_{15}^2<a_{13}a_{16}. 
\end{eqnarray}
\begin{figure}[htpb!]
    \centering
    \includegraphics[width=0.75\linewidth]{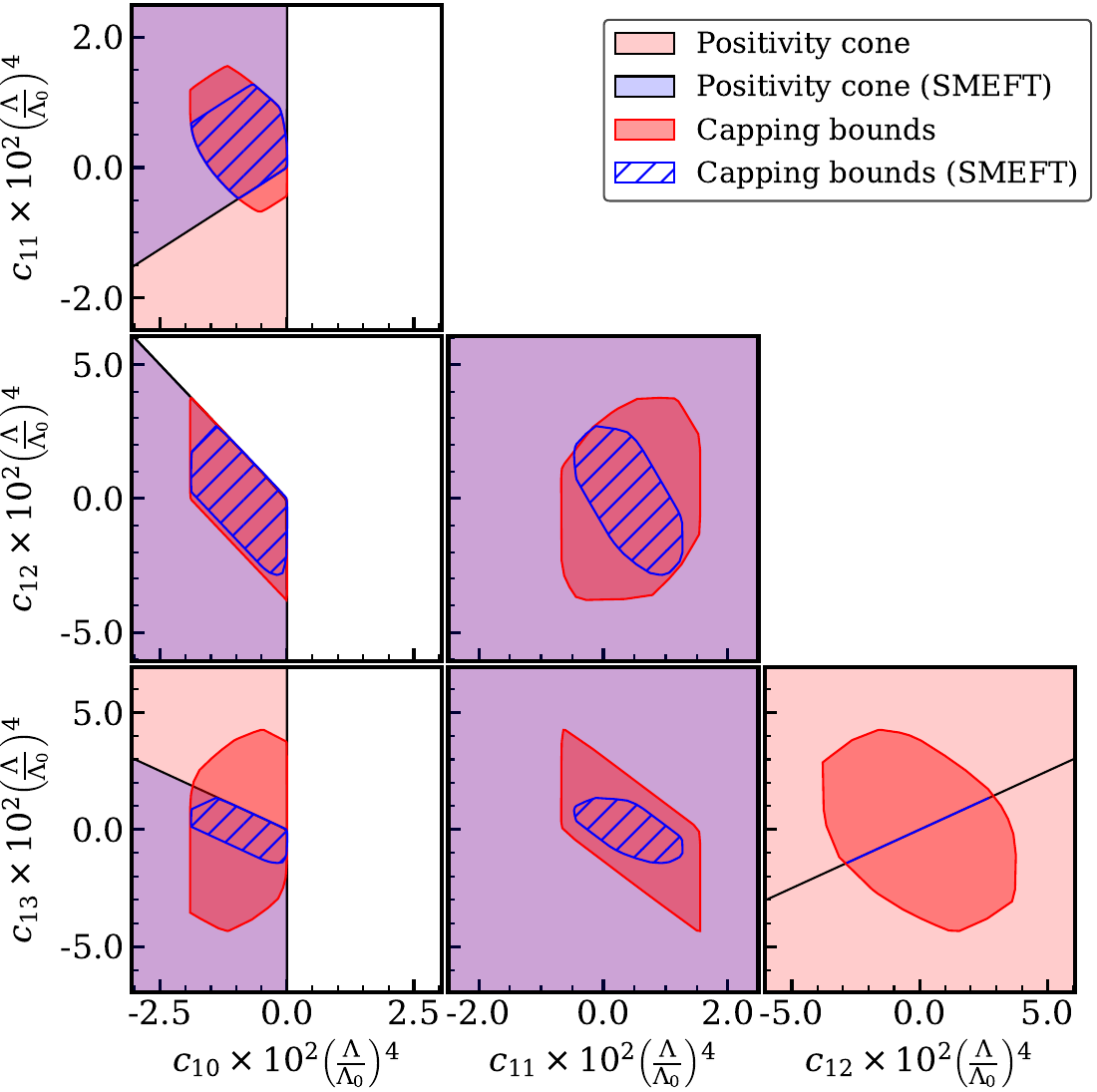}
    \caption{This figure compares the regions allowed by positivity constraints for NLO HEFT operators  contributing to longitudinal vector boson scattering, \( V_L V_L \to hh \), before and after imposing the requirements of the dimension-8 SMEFT (see Table~\ref{tab:SMEFT_constraints}). The pink region represents the   bounds for HEFT, derived using the constraints listed in Table~\ref{relations}, while the darker red region shows the capping bounds for HEFT, obtained through the numerical procedure described in Sec.~\ref{dsided}. The light blue  region (the blue hatched region) represents the regions  consistent with the analytical positivity (numerical capping) bounds as well as the SMEFT requirements of Table~\ref{tab:SMEFT_constraints}. Here, we have taken, $\Lambda_0 = 1.8$ TeV.}

    \label{fig:VVhh_smeft_vs_heft}
\end{figure}

\begin{figure}
    \centering
    \includegraphics[scale=0.5]{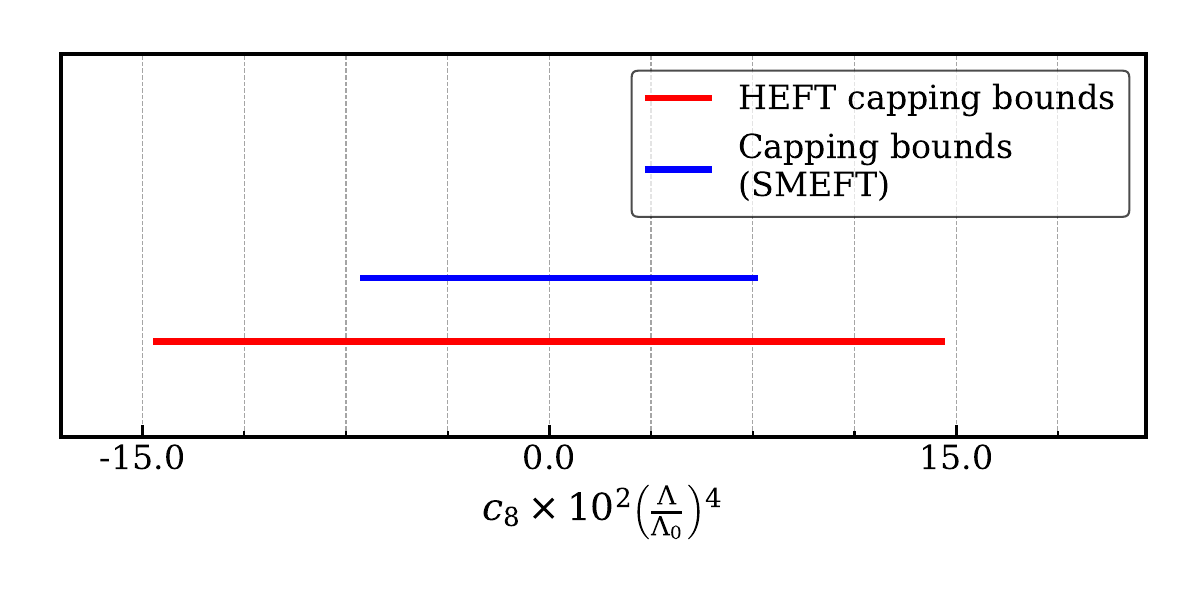}
    \caption{This figure compares the capping bounds on the WC \( c_8 \) with and without application of the SMEFT constraints. After imposing the SMEFT constraints listed in Table~\ref{tab:SMEFT_constraints}, \( c_8 \) remains the only non-vanishing WC contributing to the \( V_LV_L \to V_Lh \) process. The blue line represents the capping bound on \( c_8 \) with SMEFT constraints, while the red line shows the bound without them. The reference scale is set to \( \Lambda_0 = 1.8 \, \text{TeV} \).}

    \label{fig:VVVh_smeft_vs_heft}
\end{figure}

Next we consider all possible   $2\times2$ principle sub-matrices of $\gamma_{\bm{\beta}}$ and derive the conditions on $a_i~$s so that these sub-matrices remain positive definite irrespective of $\bm{\beta}$. We  show  explicitly the process for one of the sub-matrices. Consider the following  sub-matrix obtained from $\gamma_{\bm{\beta}}$ by deleting the second and fourth columns and rows,
\begin{eqnarray}
\begin{bmatrix}
    \left(\gamma_{\bm{\beta}}\right)_{11} & \left(\gamma_{\bm{\beta}}\right)_{13}\\
    \left(\gamma_{\bm{\beta}}\right)_{31} & \left(\gamma_{\bm{\beta}}\right)_{33}
\end{bmatrix}
\succ 0
\end{eqnarray}
We get following constraints from positive-definiteness of this matrix by marginalizing over $\bm{\beta}$,
\begin{eqnarray}
    4a_{3}^2&<&(a_6 + \sqrt{a_1a_5})^2\hspace{1.0cm}\textit{$\beta_1/ \beta_3 = (a_5/a_1)^{1/4}$ and $\beta_{2,4} = 0$}\nonumber\\
    4a_{7}^2&<&\left(\sqrt{a_6a_{11}}+\sqrt{a_1a_{13}}\right)^2\hspace{0.2cm}\textit{$\beta_1/\beta_4 = \left(\sqrt{a_{11}a_{13}}/\sqrt{a_{1}a_{6}}\right)^{1/2}$ and $\beta_{2,3} = 0$}\\
    4a_{10}^2&<&\left(\sqrt{a_6a_{11}}+\sqrt{a_4a_{13}}\right)^2\hspace{0.2cm}\textit{$\beta_2/ \beta_4 = \left(\sqrt{a_{11}a_{13}}/\sqrt{a_{4}a_{6}}\right)^{1/2}$ and $\beta_{1,3} = 0$}.\nonumber
\end{eqnarray}
\section{Implications of SMEFT constraints on the $V_L V_L \to V_L V_L, hh$ process}
\label{smeft}

In this appendix we present in  Fig.~\ref{fig:VVhh_smeft_vs_heft} (Fig.~\ref{fig:VVVh_smeft_vs_heft}) the regions in Fig.~\ref{figvvhh} (Fig.~\ref{figvvvh}) consistent with the requirements of the dimension-8 SMEFT in Table~\ref{relations}. In Fig.~\ref{fig:VVhh_smeft_vs_heft} we again  see that the region allowed by positivity is larger in the HEFT than in the SMEFT  for many of the cases. This includes the $c_{10}-c_{11}$ plane of the two custodial invariant operators.

As far as the WCs contributing to, $V_L V_L \to V_L h$, are concerned only one of them, $c_8$, is allowed to be non-zero in the dimension-8 SMEFT (see Table~\ref{relations}). We show in Fig.~\ref{fig:VVVh_smeft_vs_heft} that its allowed range shrinks significantly once we impose the SMEFT restrictions of Table~\ref{relations}.

\bibliographystyle{JHEP}
\bibliography{heft_positivity_reference.bib}
\end{document}